\documentclass[preprint]{aastex}
\newcommand{\ihmpc}{h{\rm\,Mpc}^{-1}}
\newcommand{\hmpcC}{h^{-3}{\rm\,Mpc^3}}

\newcommand{\bfw}{{\mathbf{w}}}

\newcommand{\bfP}{{\mathbf{P}}}
\newcommand{\bfPnorm}{{\mathbf{P}_{\rm norm}}}

\newcommand{\Kvect}{{\vec{K}}}
\newcommand{\xvect}{{\vec{x}}}
\newcommand{\thetavect}{{\vec{\theta}}}

\newcommand{\Pbias}{P_{\rm bias}}
\newcommand{\ra}{r_a}
\newcommand{\what}{\hat{w}}
\newcommand{\rms}{{\it rms}}

\newcommand{\Phat}{\hat{\bf P}}

\newcommand{\tableskip}{\\[-6pt]}
\newlength{\tskip}
\newlength{\colwidth}
\newcommand{\colskip}{@{\hspace{0.3in}}}
\newcommand{\etal}{et al.}

\newcommand{\beq}{\begin{equation}}
\newcommand{\eeq}{\end{equation}}
\newcommand{\beqa}{\begin{eqnarray}}
\newcommand{\eeqa}{\end{eqnarray}}

\begin{document}
\title{Correlations in the Spatial Power Spectra Inferred from Angular
Clustering: Methods and Application to APM}
\author{Daniel J. Eisenstein\altaffilmark{1,2,3} and Matias 
Zaldarriaga\altaffilmark{1,3}}
\affil{${}^1$Institute for Advanced Study, Princeton, NJ 08540\\
${}^2$Enrico Fermi Institute, 5640 South Ellis Ave., Chicago, IL 60637}

\begin{abstract}
We reconsider the inference of spatial power spectra from angular
clustering data and show how to include correlations in both the angular
correlation function and the spatial power spectrum.  Inclusion of
the full covariance matrices loosens the constraints on
large-scale structure inferred from the APM survey by over 
a factor of two.
We present a new
inversion technique based on singular value decomposition that allows one
to propagate the covariance matrix on the angular correlation function
through to that of the spatial power spectrum and to reconstruct smooth
power spectra without underestimating the errors.  
Within a parameter space of the CDM shape $\Gamma$ and the amplitude
$\sigma_8$, we find that the angular correlations in the APM survey
constrain $\Gamma$ to be 0.19--0.37 at 68\% confidence when fit to scales
larger than $k=0.2\ihmpc$.  
A downturn in power at $k<0.04\ihmpc$ is significant 
at only 1--$\sigma$.  These results are optimistic as we include only
Gaussian statistical errors and neglect any boundary effects.
\end{abstract}

\keywords{cosmology: theory -- large-scale structure of the universe}
\altaffiltext{3}{Hubble Fellow}

\section{Introduction}\label{sec:intro}
Even without distance measurements, the large-scale clustering of galaxies
can be measured through its projection on the celestial sphere.
The angular correlation function and power spectrum provide useful statistics 
to quantify
this clustering; however, in order to compare the results to theoretical models
or between different surveys, it is necessary to account for the 
projection along the line of sight \citep{Lim53,Pee73,Gro77}.
One approach to this is to 
deproject the angular statistic to the full spatial power spectrum
by assuming the latter to be isotropic in wave number.  
This inversion, however, requires some form of smoothing, which
in turn complicates the propagation of errors.  
In particular, the correlations of different scales in the angular correlation
function and the spatial power spectrum are never negligible and must be handled
correctly.

In this paper, we present improvements to two aspects of the
deprojection problem.  
First, we calculate the covariance matrix of 
the angular correlation function, and the spatial power spectrum
derived therefrom, under the approximation of wide sky coverage
and Gaussian statistics.  The former condition means that we 
neglect boundary effects; the latter condition means that we 
neglect contributions from the three- and four-point functions.
These are reasonable approximations for
the large-angle clustering signal in wide-field sky surveys
such as the Automated Plate Measuring (APM) galaxy survey \citep{Mad90}, 
Palomar Digital Sky Survey (DPOSS, Djorgovski \etal\ 1998), and 
the Sloan Digital Sky Survey (SDSS)\footnote{http://www.sdss.org/}.
We include both sample variance
and shot noise contributions, although the latter is negligible
on large angular scales in these surveys.  
While we cannot estimate the effects of systematic errors, the 
statistical covariances should provide a lower limit on the uncertainties.
We find these limits to be substantially less restrictive than
results from earlier analyses.

Second, we present a new, simple inversion technique, based on
singular value decomposition (SVD).  We use SVD to identify those
excursions in the power spectrum that would have minimal effects
on the angular clustering observables.  We then restrict these
directions from having unphysical and numerically intractable
effects on the inversion.
The errors on the observed angular correlations can be easily 
propagated to the power spectrum, including the non-trivial
correlations between different bins.  The best-fit power spectra
and covariance matrix converge as the binning in angle and 
wavenumber is refined. 

We then apply both of these improvements to the problem of inferring
the spatial power spectrum from the angular clustering
of the APM galaxy survey \citep{Mad90}.  
Assuming only Gaussian statistical errors, we reconstruct the binned 
bandpowers and their covariance matrix.  We find large anti-correlated 
errors.  To give a sense of what these results imply for the 
measurement of the power spectrum on large scales, we fit scale-invariant
CDM models to the results at $k<0.2\ihmpc$.  
Varying only the shape parameter $\Gamma$ and the primordial amplitude,  
we find that $\Gamma$ is constrained to be 0.19--0.37 (68\%).
Inclusion of non-Gaussianity, survey boundary effects, or systematic
errors  could make this constraint weaker.
While CDM models with $\Gamma\approx0.25$ are good fits to the data,
it is important
to note that the statistical power of this fit is dominated by $k>0.1\ihmpc$.
A turnover in power at $k<0.04\ihmpc$ is detected at only 1--$\sigma$.
We would therefore not say that the large-angle clustering of APM
confirms the shape of the CDM model power spectrum.

One could get tighter limits on $\Gamma$ by extending the fit to smaller
scales.  
On small scales, however, scale-dependent bias and 
non-linear evolution may minimize and obscure the differences between
cosmologies.  It is on large scales where the details of the
shape of the power spectrum draw unambiguous distinctions 
between cosmological models.  Hence, we have a particular interest
in what can be learned at large scales.

Our large-scale constraints are over a factor of two looser than 
earlier results in the literature.
\citet[hereafter BE93; 1994]{Bau93} used Lucy inversion to infer
the spatial power spectrum from the APM angular correlation function
\citep{Mad96}.  The errors on this power spectrum could only be estimated
as the deviation between 4 subsamples of the survey.  The small number of
subsamples prevented an estimation of the covariance between different
wavenumber bins.  It is clear that correlations between the subsamples
are non-negligible.  Moreover, because the smoothing of the power
spectra in each subsample was done before the dispersion was computed,
the errors are substantially underestimated in poorly constrained regions.
Both of these effects lead to overly optimistic constraints.  Recently,
\citet[hereafter DG99]{Dod99} presented a different inversion technique
based on a Bayesian smoothness prior.  Their method allows one to estimate
the covariance of the spatial power spectrum from the covariance on the
angular correlation function.  However, they include only the diagonal
elements of the latter covariance matrix.  We show that this leads to a
factor of two underestimate of the error bars on CDM model parameters.
Moreover, like the BE93 method, the DG99 technique systematically
underestimates error bars in poorly constrained regions.

The structure of this paper is as follows.
In \S\ \ref{sec:defns}, we present the definitions for clustering
statistics and the relations between them.  In \S\ \ref{sec:cov},
we show how to calculate the covariance matrix for the angular
correlation function.  \S\ \ref{sec:svd} describes how to 
construct the spatial power spectrum using SVD.  We then apply 
these methods to the APM angular clustering in \S\ \ref{sec:apm},
recovering the correlated bandpowers in \S\ \ref{sec:apm_bandpow}
and fitting them to CDM models in \S\ \ref{sec:cdm}.
In \S\ \ref{sec:nongauss},
we consider the effects of non-Gaussianity
and estimate that they are likely to be small.
In \S\ \ref{sec:best}, we demonstrate that the constraints obtained
are close to the best-possible errors available to an angular
clustering survey with the selection function and sky coverage of APM.
We compare our work to previous analyses in \S\ \ref{sec:diag}.
We conclude in \S\ \ref{sec:concl}.

\section{Definitions and Relations}\label{sec:defns}

Following the usual notation, we take the angular positions of
the galaxies to define a continuous fractional overdensity field
$\delta(\xvect)$, where $\xvect$ is a position on the sky.
We take a flat-sky approximation and define the Fourier modes of
this density field as 
$\delta_\Kvect = \int d^2x \delta(\xvect) e^{-i\Kvect\cdot \xvect}$
for all angular wavevectors $\Kvect$.
If the random process underlying the density field is 
translationally-invariant, then ensemble averages of the product
of two of these Fourier modes is given by the power spectrum:
\beq\label{eq:Pdef}
\left<\delta_\Kvect \delta^*_{\Kvect'}\right>
= (2\pi)^2 \delta_D^{(2)}(\Kvect-\Kvect') P_2(\Kvect).
\eeq
$\delta_D^{(2)}$ is the two-dimensional Dirac delta function.
The power spectrum $P_2$ is the sum of the true power spectrum
and a shot noise term equal to the inverse of the number density
of sources on the sky.
We will assume that $P_2$ is isotropic.
The angular correlation function is defined as
\beq\label{eq:wPft}
w(\theta) \equiv \left<\delta(\xvect)\delta(\xvect+\thetavect)\right>_\xvect = 
\int {d^2K\over(2\pi)^2} e^{i\Kvect\cdot\thetavect} P_2(K)
= \int {K\,dK\over 2\pi} J_0(K\theta) P_2(K),
\eeq
where $J_0(x)$ is the Bessel function.

Relating these angular correlations to their parent three-dimensional
correlations requires one to include the survey-dependent projection
along the line-of-sight.  We adopt the Limber approximation to 
project the spatial clustering 
\citep{Lim53,Gro77,Phi78}.  
This is valid for modes with wavelengths smaller
than the survey depth and any evolutionary scale.

The projection is characterized by the redshift distribution $dN/dz$ 
of the galaxies in the survey.  The total number of galaxies per unit
solid angle is denoted $N$.  The cosmology and the 
evolution of clustering affect the projection, although for analysis
of APM, the differences can be scaled out easily.
As we are interested in
large scales, we assume that the power spectrum can be separated into
a function of redshift $z$ and a function of {\it comoving} spatial 
wavenumber $k$
\beq
P(k,t) = {P(k)\over(1+z)^a},
\eeq
where $P(k)$ denotes the present-day spatial power spectrum (BE93).
The function of time is a convolution of the growth of perturbations in the
mass, the time evolution of bias, and the effects of 
luminosity-dependent bias between nearby, faint galaxies and distant,
bright ones.  
Following the notation of BE93, the angular power spectrum is
\beq\label{eq:Plimber}
P_2(K) = {1\over K} \int{dk P(k) f(K/k)}
\eeq
where the kernel is
\beq
f(r_a) = \left[{1\over N}{dN\over dz}{dz\over d\ra}\right]^2
	{F(\ra) \over (1+z)^a}.
\eeq
Here, $\ra=K/k$ is the comoving angular diameter distance
(or proper motion distance) to a redshift $z$.
One has the simple relation
\beq
{dz\over d\ra} = E(z) = 
	\left[\Omega_m(1+z)^3+\Omega_K(1+z)^2+\Omega_\Lambda\right]^{1/2}
\eeq
where $\Omega_m$ is the density in non-relativistic matter, 
$\Omega_\Lambda$ is the cosmological constant, and
$\Omega_K=1-\Omega_m-\Omega_\Lambda$.
Curvature also enters through the volume correction
\beq
F(\ra)=\sqrt{1+(H_0\ra/c)^2\Omega_K}.
\eeq
Combining equations (\ref{eq:wPft}) and (\ref{eq:Plimber}), 
we can write the angular correlation function as
\beqa\label{eq:wlimber}
w(\theta) &=& \int_0^\infty kP(k) g(k\theta) dk \\
g(k\theta) &=& {1\over 2\pi} \int d\ra  J_0(k\theta\ra)
{F(\ra)\over (1+z)^a} \left[{1\over N}{dN\over dz}{dz\over d\ra}\right]^2.
\eeqa

\section{Covariance}\label{sec:cov}
We are interested in the estimation of the angular correlation function
on large angular scales in wide-field surveys.  
In these surveys, the density of galaxies is large enough that including only
shot noise---the sparse sampling of the density field by 
the galaxies---would severely underestimate the errors.  
Instead, the errors are dominated by ``sample variance'', the
uncertainty due to the finite number of 
patches of the desired angular scale available within the survey.
If the angular extent of the survey is large 
compared to the correlation scales and compared to the angular projection
of any clustering scales, then corrections from the boundaries of the
survey will be small.
In this limit, the effects of sample variance on the angular correlation
function can be easily calculated.

Imagine that our survey has a selection window $W(\xvect)$ on the sky,
with $W=1$ in covered regions and $W=0$ elsewhere.  Then the 
estimator of $w(\thetavect)$ is simply
\beqa
\what(\thetavect) &=& {1\over A(\thetavect)} 
\int d^2x W(\xvect) \int d^2x' W(\xvect')
\delta(\xvect) \delta(\xvect') \delta_D^{(2)}(\xvect-\xvect'-\thetavect) \\
&=& \int {d^2K\over(2\pi)^2}{d^2K_1\over(2\pi)^2}
\delta_\Kvect \delta^*_{\Kvect_1} e^{i\Kvect_1\cdot\thetavect}
h(\Kvect-\Kvect_1, \thetavect)
\eeqa
where
\beq
A(\thetavect) = \int d^2x W(\xvect) W(\xvect+\thetavect)
\eeq
and 
\beq
h(\Kvect, \thetavect) = {1\over A(\thetavect)} \int d^2x 
e^{i\Kvect\cdot\thetavect} W(\xvect) W(\xvect+\thetavect).
\eeq
Using Equations (\ref{eq:Pdef}) and (\ref{eq:wPft}), 
one finds that $\left<\what(\thetavect)\right>=w(\thetavect)$.

The covariance of this set of estimators can be written
\beqa\label{eq:Cwfull}
C_w(\thetavect,\thetavect') &\equiv&
\left<[\what(\thetavect)-w(\thetavect)]
[\what(\thetavect')-w(\thetavect)]\right> \\
&=& \int {d^2K\over(2\pi)^2} {d^2K_1\over(2\pi)^2}
e^{i\Kvect_1\cdot\thetavect} h(\Kvect - \Kvect_1, \thetavect)
\int {d^2K'\over(2\pi)^2} {d^2K'_1\over(2\pi)^2}
e^{i\Kvect'_1\cdot\thetavect'} h(\Kvect'-\Kvect'_1, \thetavect') \\
&& \times \left[\left<\delta_\Kvect \delta^*_{\Kvect_1}
\delta_{\Kvect'} \delta^*_{\Kvect'_1}\right> -
\left<\delta_\Kvect \delta^*_{\Kvect_1}\right>
\left<\delta_{\Kvect'} \delta^*_{\Kvect'_1}\right> \right].
\eeqa
The expectation of four $\delta$'s involves a Gaussian term as well
as the four-point function:
\beq\begin{array}{l}
\left<\delta_\Kvect \delta^*_{\Kvect_1}
\delta_{\Kvect'} \delta^*_{\Kvect'_1}\right> -
\left<\delta_\Kvect \delta^*_{\Kvect_1}\right>
\left<\delta_{\Kvect'} \delta^*_{\Kvect'_1}\right> = \\[3pt]
\hspace{1in}(2\pi)^2\delta_D^{(2)}(\Kvect+\Kvect')P_2(\Kvect)
(2\pi)^2\delta_D^{(2)}(\Kvect_1+\Kvect_1')P_2(\Kvect_1) + \\[3pt]
\hspace{1in}(2\pi)^2\delta_D^{(2)}(\Kvect-\Kvect_1')P_2(\Kvect)
(2\pi)^2\delta_D^{(2)}(\Kvect_1-\Kvect')P_2(\Kvect_1) + \\[3pt]
\hspace{1in}(2\pi)^2\delta_D^{(2)}(\Kvect-\Kvect_1+\Kvect'-\Kvect_1)
T_4(\Kvect,\Kvect_1,\Kvect',\Kvect_1').
\end{array}
\eeq
The four-point function $T_4$ primarily includes the four-point
function of the density, but non-zero shot-noise also introduces
terms involving the two- and three-point functions \citep{Ham99}.
We can simplify the Gaussian portion of $C_w(\thetavect,\thetavect')$ to
\beq\label{eq:Cwgauss}
C_w(\thetavect,\thetavect') 
= \int {d^2K\over(2\pi)^2} {d^2K_1\over(2\pi)^2} P_2(\Kvect) P_2(\Kvect_1) 
h(\Kvect-\Kvect_1, \thetavect) h^*(\Kvect-\Kvect_1, \thetavect')
\left(e^{i\Kvect_1\cdot\thetavect+i\Kvect\cdot\thetavect'} +
e^{i\Kvect_1\cdot\thetavect-i\Kvect_1\cdot\thetavect'}\right).
\eeq

We will drop the non-Gaussian terms from equation (\ref{eq:Cwfull}).
If one assumes Gaussian initial perturbations, then one can use the
Gaussian analysis to study large angular and spatial scales.  On smaller
scales, we expect that non-Gaussianity will contribute considerably more
variance due to the correlations between modes \citep{Mei99,Sco99}.  
It is important to note
that angular clustering statistics tend to have smaller non-Gaussian 
terms than a
simple mapping of the spatial non-linear scale would suggest.  This is
because one is projecting many non-linear regions along the line-of-sight;
the central limit theorem then drives the sum of the fluctuations towards
Gaussianity.  We note that although this is comforting for the calculation
of the angular correlations, it is not clear that the inference of
spatial clustering from the angular statistics retains this advantage.

For the particular case of APM, calculations with the hierarchical
ansatz (\S~\ref{sec:nongauss}) 
suggest that non-Gaussian terms become equal to the Gaussian
terms at $K\gtrsim100$, which indicates that our smallest scales are
not safely in the Gaussian regime.  Unfortunately, the ansatz is not
reliable enough to give a useful calculation of the 4-point terms in
equation (\ref{eq:Cwfull}) \citep{Sco99}.  
As we will describe in \S\ \ref{sec:nongauss}, a
simple attempt to include non-Gaussianity degraded 
our results by $\sim$10\%.
We therefore regard non-Gaussianity as a caveat to our results but not
a catastrophic error.

We are interested in the case when the sky coverage of the survey
is large, both compared to the angle $\theta$ and to any angular
correlation length.  Here, the function $h(\Kvect,\thetavect)$ becomes
sharply peaked around $\Kvect=0$.  To leading order, it may be treated
as a Dirac delta function.  The coefficient is
\beq
\int {d^2K\over (2\pi)^2} h(\Kvect,\thetavect) h^*(\Kvect,\thetavect')
= {\int d^2x W^2(\xvect) W(\xvect+\thetavect) W(\xvect+\thetavect')\over
A(\thetavect) A(\thetavect')} = {1\over A_\Omega},
\eeq
where $A_\Omega$ is simply the area of the survey.
Effects from boundaries or from features in the power spectrum will
be suppressed by another power of $A_\Omega$.
For wide-angle surveys, we can approximate the correlations in $w(\theta)$
as 
\beq
C_w(\thetavect,\thetavect') = {1\over A_\Omega} \int {d^2K\over(2\pi)^2} 
P_2^2(\Kvect) \left[e^{i\Kvect\cdot(\thetavect+\thetavect')} +
e^{i\Kvect\cdot(\thetavect-\thetavect')} \right].
\eeq
Since we are neglecting all boundary effects, we can average $\thetavect$
over angle, i.e.\ $w(\theta)=(1/2\pi)\int d\phi\,w(\thetavect)$, to yield
\beq\label{eq:Cww}
C_w(\theta,\theta') \equiv 
\left<[\what(\theta)-w(\theta)][\what(\theta')-w(\theta')]\right> =
{1\over\pi A_\Omega} \int_0^\infty dK\,K P_2^2(K) J_0(K\theta) J_0(K\theta').
\eeq
This is the Gaussian contribution to the covariance 
of the angular correlation function in the limit of a wide-field survey.
Our neglect of the boundary terms is equivalent to the approximation
that working on a fraction $f_{\rm sky}$
of the sky simply increases the variance on the angular power spectra by 
$f_{\rm sky}^{-1}$ \citep{Sco94}.

It is important to note that as the area of a survey increases
the covariance in equation \ref{eq:Cww} {\it does not} approach 
a limit in which errors 
on different angular scales are statistically independent.  
This means that analysis of $w(\theta)$ in the
sample-variance limit must not neglect the correlation of the error
bars on $w(\theta)$.  This runs contrary to the properties of 
$P_2$, in which differing scales do become independent in the
large-data-set limit of a Gaussian process.  We will show later that
the inclusion of these correlations substantially weakens the 
published constraints on the large-scale power spectrum from the APM survey.

Our estimate of $C_w$ does not include systematic errors,
the effects of non-Gaussian statistics, or aliasing from 
the survey boundary.  
It would be very surprising, 
however, if these complications were to reduce the uncertainty on
inferring $P(k)$!  In this sense, we consider Equation (\ref{eq:Cww})
as a lower bound on the errors.

\section{SVD Inversion}\label{sec:svd}

We wish to estimate $P(k)$ from observations of $w(\theta)$.  In
practice, we are given estimates of $w$ in $N_\theta$ bins centered on
angles $\theta_j$ ($j=1,\ldots,N_\theta$).  We denote the estimates
as $w_j$ and place them in a vector $\bfw$.  These measurements have
a $N_\theta\times N_\theta$ covariance matrix $C_w$.  

We then wish to estimate $P(k)$ in $N_k$ bins centered at $k_j$ 
($j=1,\ldots,N_k$).  The values in these bins are denoted $P_j$
and formed into a vector $\bfP$.  The integral transform of 
equation (\ref{eq:wlimber}) can then be cast as a matrix, yielding
\beq\label{eq:discrete}
\bfw = G \bfP,
\eeq
where $G$ is a $N_\theta\times N_k$ matrix.

In detail, one should calculate the elements of $G$ taking account of
the averaging in the bins of $k$ and $\theta$.  The method of averaging
can be chosen, but if one takes the estimates $w_j$ to be averages
of $w(\theta)$ according to the weight $\theta d\theta$ and treats
$P(k)$ as constant within a bin in $k$, then the integrals over $k$
and $\theta$ can be done analytically using properties of the Bessel
function.  We find, however, that for reasonably narrow bins
the approximate treatment of using only the central values of $\theta$
and $k$ produces nearly the same answer as the exact integration.

With this notation, the best-fit power spectrum is simply\footnote{if $G$
were square; if not, the formula is $\Phat= C_P G^T C_w^{-1} \bfw$,
exactly as one would get from SVD.}
$\Phat = G^{-1} \bfw$ and the covariance matrix on this inversion
is $C_P^{-1} = G^T C_w^{-1} G$.  It is important to note that this
covariance matrix is {\it not} diagonal, even in the large survey
volume limit.  This differs from the behavior of estimates of $P_3(k)$
from a redshift survey, where individual bins approach independence in
the large-volume limit.

In practice, the matrix $G$ is nearly singular, as one would guess
from its origin as a projection from 3 dimensions to 2.  By singular,
we mean that the matrix has a non-zero null space, i.e.\ that there
are vectors $\bfP$ that are annihilated by $G$.  Such null directions
cannot be constrained from the angular data.  Often the near 
singularities come about from having too fine a binning in $k$ or from
extending the domain in $k$ to values that have negligible impact on
the range of angular scales being measured.  Left untreated, these
directions introduce wild excursions in $P(k)$ in order to compensate tiny
variations in $w(\theta)$.  The resulting covariance matrix $C_P$ 
has enormous, but highly anti-correlated, errors.

Singular value decomposition offers a useful way to treat this singularity.
If we take the measurements $\bfw$ to be Gaussian-distributed 
around their true values $\bfw_m$, then the distribution of the observations
follows the probability $P\propto \exp(-\chi^2/2)$, where
\beq
\chi^2 = (\bfw_m-\bfw)^T C_w^{-1} (\bfw_m-\bfw).
\eeq
$\bfw_m$ is related to the true power spectrum by $\bfw_m = G \bfP_m$.
We will rescale the basis set for $\bfP_m$ by dividing by a set of
reference values $\bfPnorm$, intended to be defined by a fiducial
power spectrum $P_{\rm norm}(k)$ evaluated at the appropriate wavenumbers.
This produces $\bfP'$ by dividing each element of $\bfP$ by the 
corresponding element of $\bfPnorm$.
We also define $\bfw' = C_w^{-1/2} \bfw$ and 
$G' = C_w^{-1/2} G \bfPnorm$; here, $C_w^{-1/2}$ is constructed by taking the
inverse square root of the eigenvalues of the positive-definite
$C_w$ matrix.  
We then have
\beq\label{eq:chi}
\chi^2 = |G'\bfP'_m-\bfw'|^2.
\eeq
Finding the vector (or subspace) $\bfP'_m$ that minimizes $\chi^2$,
and thereby 
maximizes the likelihood in a Gaussian treatment, is a prime application
of SVD, and the technique allows one to treat the nearly singular
directions in $P$-space explicitly.  Note that we can immediately see
that the covariance matrix of the $\bfP'_m$ will simply be $(G'^T G')^{-1}$.
We will now drop the $m$ subscript and refer to the reconstructed 
power spectrum as $\bfP'$.

We define the SVD of the $G'$ matrix by $G'=UWV^T$ 
\citep[for a review]{Pre92}, where
$W$ is a square, diagonal matrix of the singular values (SV), $V$ is a
$N_k \times N_k$ orthogonal matrix, and $U$ is a $N_\theta\times N_k$
column-orthonormal matrix.  Singular values close to zero correspond
to columns in $V$ that contain $\bfP'$ directions that have almost
no effect on $\bfw'$ and therefore are not well-constrained.
To find the best-fit power spectrum, we use 
$\bfP' = V W^{-1} U^T \bfw'$.
The covariance matrix $C_P$ of $\bfP'$ is simply $V W^{-2} V^T$, 
which is the diagonalization of the covariance matrix.

Mathematically, the fact that the $W$ matrix is diagonal means that
the data $\bfw'$ is coupled to the power spectrum estimate $\bfP'$
through $N_k$ distinct modes, in which the matching columns of the
$U$ and $V$ matrix specify a matching set of $\bfw'$ and $\bfP'$
excursions.
We denote the $j$th SV as $W_j$ and the $j$th 
column of $U$ and $V$ as $U_j$ and $V_j$, respectively.
We refer to the set $W_j$, $U_j$, and $V_j$ as the $j$th SV mode.
In detail, each mode enters the best-fit power spectrum as
$P_i = P_{{\rm norm},i} V_{ji} W_j^{-1} U_j^T \bfw'$.
Comparing this to the formula for $C_P$ shows that 
$U_j^T\bfw'$ is the number of standard deviations by which 
the $j$th mode is demanded by $\bfw'$.   
Indeed, $\chi^2$ may be rewritten as
\beq
\chi^2 = |\bfw'|^2 - \sum_j |U_j^T \bfw'|^2,
\eeq
so the value of $(U_j^T\bfw')^2$ is the amount by which the
inclusion of a mode in $\bfP'$ will decrease $\chi^2$.  
Since the columns of $V$ are unit-normalized, the quantity 
$W_j^{-1} U_j^T \bfw'$ is a measure of the size of the contribution
that this mode makes to the power spectrum in units of $\bfPnorm$.

Of course, such an analysis is only useful if it converges as
the binning of $k$ and $\theta$ becomes finer.
In our APM example (\S\ \ref{sec:apm}), we find that this is the case:
the columns in $U$ and $V$ corresponding to large singular values
change very little as we alter the binning in $k$ or $\theta$.
The large $W_j$ scale as $N_k^{-1/2}$
because if one simply refines the binning in $k$, the elements of
$G$ scale as $N_k^{-1}$ due to the smaller range $dk$ in the defining
integral while 
the elements of $V_j$ scale as $N_k^{-1/2}$ because it is an unit-normalized
vector.
The primary effect of adding or removing bins is to change 
the number of tiny singular values.
The modes with large SV show broad tilts and curves in $P(k)$;
the modes with small SV show rapid compensating oscillations as well as
excursions at very large or small $k$ that the $w(\theta)$ data
don't constrain.  The ability to identify the
excursions in $P(k)$-space that are well-constrained, in a 
manner that converges as the binning becomes finer, is the 
strength of the SVD method.
It should be noted that the kernel and its SVD decomposition depend on
the survey geometry, the $C_w$ covariance matrix, and the fiducial 
scaling $\bfPnorm$, but not upon the observed data $\bfw$ itself.

Left untreated, the small singular values will have large inverses
and therefore produce large excursions in $\bfP'$.  Such excursions
are unphysical and can even make $C_P$ numerically intractable.
In the usual spirit of SVD, we wish to adjust the treatment of these singular
values.
This is complicated by the fact that SVD relies upon the concepts 
of orthogonality and normalization and thereby
implies a geometric structure that our $P$- and $w$-spaces don't actually
have.  To sort the singular values and declare some of them to be ``small''
requires that we have some sense of comparing $w$ at different values of
$\theta$ or $P$ at different values of $k$.  
On the $\bfw'$-space side, this choice is easy: 
the absorption of $C_w$ into $G'$ and $w'$ means that the unit-normalization
of fluctuations in $\bfw'$ have the correct role in the $\chi^2$ statistic.
However, for $\bfP'$-space, the choice is more arbitrary.
The fiducial power spectrum $\bfPnorm$ determines how fluctuations
in power on different scales but of equal statistical significance
are to be weighted in the singular values.  By choosing $\bfPnorm$
to be close to the observed spectrum, we are opting that equal
fractional excursions on different scales receive equal weight.
Had we instead chosen $\bfPnorm$ to be a constant, a 100\% oscillation
at the peak of the power spectrum ($P\approx 10^4\hmpcC$ at 
$k\approx0.05\ihmpc$) 
would have been suppressed relative to the same fractional fluctuation 
at smaller scales (say, $P\approx 10^2\hmpcC$ at $k\approx1\ihmpc$).
It is important to remember that $\bfPnorm$ is irrelevant if one
is using all of the singular values unmodified.  It enters only
when we place a threshold on the singular values (as described below).
When small singular values are altered or eliminated, $\bfPnorm$
determines how different scales are to be compared in the application
of a smoothness condition.
While the arbitrary choice of $\bfPnorm$ means that an SVD treatment 
of the inversion is not unique, we feel that our $\bfPnorm$ is a well-motivated choice:
each singular value represents the square root of the
$\chi^2$ contribution for a given fractional excursion around the 
best-fit power spectrum.  

We next describe how we alter the small $W_j$.
In detail, we incorporate two different SV thresholds, one for the 
construction of $C_P$ and another for the construction of $\bfP'$.
Small SV indicate poorly constrained directions.  We would like
$C_P$ to reflect this, but not to the extent that the matrix
becomes numerically intractable.  Physically, these small $W_j$ are
highly oscillatory, and our prior from both theory and previous
observations is that enormous (much greater than unity) fluctuations
don't exist in the power spectrum.  Hence, when constructing $C_P$,
we increase all SV to a minimum level of $SV_C$.  We can't recommend
a choice of $SV_C$ for arbitrary applications because all the $W_j$
will scale with the normalization of $\bfPnorm$.  However, in
our work, where $\bfPnorm$ has a similar amplitude to the actual
power spectrum, a choice of $SV_C$ between 0.1 and 1.0 will allow 
order unity fluctuations in the power spectrum.  This is larger than 
any oscillations ever seen but not so large as to make $C_P$ overly singular.

Without correcting the small $W_j$, the best-fit power spectrum $\bfP'$
becomes wildly fluctuating.  If we have modified the small $W_j$ in $C_P$,
then these fluctuations will appear to be highly significant.  
Hence, at a minimum one must use the same $W_j$ in $\bfP'$ as 
those used in $C_P$, so as to keep the fluctuations and the covariance
on the same scale.  However, since the small SV modes have already been
granted an error budget larger than what is likely observable, there
is no reason to include them in the best-fit $\bfP'$ at all.  The
difference between the best-fit power spectrum and any reasonably
smooth model power spectrum will be insignificant with respect to 
the covariance $C_P$.
Hence, we generally only include in $\bfP'$ the modes with the 
largest SV.  Essentially this is a threshold on $W_j$ for inclusion
in the nominal best-fit.  In practice, when comparing between 
spectra calculated with different $C_w$ (as we will occasionally do in
next section), it is better to keep a fixed number of SV modes rather
than fixed $W_j$ threshold because the $W_j$ will change even while
the SV spectrum and the structure of the $U$ and $V$ matrices 
remains fairly constant.

The modification of $W^{-1}$ when calculating
$\bfP'$ causes $\bfP'$ to be a biased estimator of the power spectrum.
This bias is statistically significant at a level of 
$(1-W_j/W_{j,\rm used})\left|U_j^T\bfw'\right|$,
where $W_{j,\rm used}$ is the value of $W_j$ actually used in 
constructing $\bfP'$ ($\infty$ if the mode has been dropped).
One can thereby judge the statistically significance of the bias 
imposed by altering a $W_j$ and decide whether an excursion of
the amplitude implied by the original $W_j$ is physically reasonable.
Of course, one only wishes to drop modes that are statistically
irrelevant or physically unreasonable.
It is important to remember that the small $W_j$ can be strongly
perturbed by small changes in $C_w$.  Since one cannot hope to control
all systematic errors in $C_w$, one doesn't want to place any 
weight on the SV modes with small $W_j$.

The bias from increasing $W_j$ or omitting modes in $\bfP'$ pulls the 
amplitude of the altered modes toward zero.  Usually
this pulls the power toward zero, but we can alter this behavior by
subtracting $G\Pbias$ from the $w(\theta)$ data and adding $\Pbias$
to the reconstructed power spectrum.  In other words, one alters
equation (\ref{eq:chi}) to
\beq
\chi^2 = |G'(\bfP'_m-\Pbias)-(\bfw'-G'\Pbias)|^2,
\eeq
and reconstructs $\bfP'_m-\Pbias$.  The bias then pulls toward $\Pbias$.

All of the above techniques apply equally well to the problem of inferring
the spatial power spectrum from the angular power spectrum, and it is
trivial to alter the equations.

\section{APM}\label{sec:apm}
\subsection{Reconstructing the Power Spectrum}\label{sec:apm_bandpow}

One of the most influential uses of angular clustering in the last
decade has been its application to the APM survey \citep{Mad90}.
However, a full treatment of the 
covariance matrix on power spectra inferred from
APM clustering has not yet been presented, and so we choose this as our
example.
We take the data on the APM angular correlation function \citep{Mad96}
as presented in the binned results of 
DG99.  This includes 40 half-degree bins from $0^\circ.5$ to $20^\circ$.  
Tests show that including data from smaller angular scales does not 
affect our results on scales larger than $k=0.2\ihmpc$.

We discard the quoted errors on the DG99 $w(\theta)$ data
and instead use the covariance matrix from \S\ \ref{sec:cov}.  
For $P_2$, we use 
\beq\label{eq:P2fit}
P_2(K) = \left\{\begin{array}{ll}
2\times10^{-4} & K<20 \\
2\times10^{-4}(K/20)^{-1.35}&K>20,
\end{array}\right.
\eeq
which is a reasonable fit to the results of \citet{Bau94}.
We assume a survey area of 1.31 steradians.
We add a shot noise term of $P_{\rm shot} = 1/\bar{n} = 10^{-6}$, 
based on a number density $\bar{n}$ of 1 galaxy per $3'.5$ 
square pixel \citep{Bau94}. 
The shot noise has little impact on the results.  
We use this $P_2$ to calculate $C_w$ according to equation (\ref{eq:Cww}).

We assume $\Omega_m=1$, $\Omega_\Lambda=0$, and $a=0$ in the calculation.
The results do depend on these choices, but nearly
all of the behavior can be scaled out in two easy parts.
First, the fact that the average galaxy in the survey is at $z\approx0.11$
means that specifying a time evolution of the power spectrum ($a\ne0$)
will cause a shift in the amplitude of the $z=0$ power spectrum.
In practice, one can consider the reconstructed power spectrum
to be appropriate to $z=0.11$; in other words, choosing different
time dependences for the power spectrum leaves the power at $z=0.11$
essentially constant.  Second, the cosmology enters through the volume
available at higher redshift.  Models with more volume per unit 
redshift (lower $dz/d\ra$) have more modes and therefore suffer 
more dilution in angular clustering when projected.  This effect scales 
roughly as $E(z=0.11)^{-2}$.  Despite the low median redshift, 
this is not a small effect in $\Lambda$ models: using an $\Omega_m=0.3$, 
$\Omega_\Lambda=0.7$ model yields a $P_3(k)$
20\% higher than our fiducial model.  The effect is
much smaller in open models.  In detail, $a\ne0$ or a change in
the cosmology will also shift the average depth of the survey
slightly, causing the reconstructed power spectrum 
to move in wavenumber.  We find this effect to be less than 5\%,
which is certainly within the errors.

We use a logarithmic binning in $k$-space.  Our fiducial set uses
3 bins per octave, ranging from $k=0.0125\ihmpc$ to $k=0.8\ihmpc$.
There is also a large-scale bin of $k<0.0125\ihmpc$, for a total
of 19 bins.  Wavenumbers greater than $0.8\ihmpc$ are not constrained
by data at $\theta>0^\circ.5$ and would simply be degenerate with
our last bin.  We also tried coarser and finer binnings, using 2 and 4
bins per octave to get 13 and 25 bins, respectively.  
These three choices are shown in Table \ref{tab:binning} under the
names $k13$, $k19$, and $k25$.  We find equivalent results with 
non-logarithmic binning schemes.

We set $\bfPnorm$ to be 
\beq
P_{\rm norm}(k) = {1.5\times 10^4 h^{-3} Mpc^3 \over 
\left[1+(k/0.05h Mpc^{-1})^2\right]^{0.65}}
\eeq
This is a rough fit to the observed power spectrum until 
$k\approx0.05\ihmpc$ and has constant power on the largest scales.  
Recall that $\bfPnorm$ enters only in the treatment of small singular
values and serves to set an upper bound on the allowed size of 
fluctuations in the power spectrum relative to the best-fit.  
The constant power on large scales was chosen so as not to prejudice the 
results on scales where we have little information.
It is important to choose $P_{\rm norm}$ to be continuous because the
prior against rapid oscillations acts on the ratio of the fitted
power spectrum to $\bfPnorm$.  Discontinuities in $P_{\rm norm}$
would impose discontinuities in the fitted power spectrum.

\begin{table}[tb]\footnotesize
\caption{\label{tab:binning}}
\begin{center}
{\sc Singular Values and $k$-space Bins\\}
\begin{tabular}{ccc\colskip ccc}
\tableskip\tableline\tableline\tableskip
\multicolumn{3}{c\colskip }{$k$ bins ($\ihmpc$)}&\multicolumn{3}{c}{Singular Values, scaled to 19 bins} \\
k13 & k19 & k25 & 
$W_{k13}\times\left({13\over19}\right)^{1/2}$ & 
$W_{k19}$ & 
$W_{k25}\times\left({25\over19}\right)^{1/2}$ \\
\tableskip\tableline\tableskip
0.000--0.0125 & 0.000--0.0125 & 0.000--0.0125 & 17.1 & 17.1 & 17.1 \\
0.0125--0.018 & 0.0125--0.016 & 0.0125--0.015 & 9.2 & 9.3 & 9.3 \\
0.018--0.025 & 0.016--0.020 & 0.015--0.018 & 5.2 & 5.3 & 5.3 \\
0.025--0.035 & 0.020--0.025 & 0.018--0.021 & 3.1 & 3.2 & 3.2 \\
0.035--0.050 & 0.025--0.032 & 0.021--0.025 & 2.0 & 2.0 & 2.1 \\
0.050--0.071 & 0.031--0.040 & 0.025--0.030 & 1.4 & 1.5 & 1.5 \\
0.071--0.100 & 0.040--0.050 & 0.030--0.035 & 1.0 & 1.1 & 1.2 \\
0.100--0.141 & 0.050--0.063 & 0.035--0.042 & 0.80 & 0.88 & 0.93 \\
0.141--0.200 & 0.063--0.079 & 0.042--0.050 & 0.53 & 0.59 & 0.62 \\
0.200--0.283 & 0.079--0.100 & 0.050--0.059 & 0.30 & 0.37 & 0.39 \\
0.283--0.400 & 0.100--0.126 & 0.059--0.071 & 0.12 & 0.22 & 0.23 \\
0.400--0.566 & 0.126--0.159 & 0.071--0.084 & 0.014 & 0.12 & 0.13 \\
0.566--0.800 & 0.159--0.200 & 0.084--0.100 & $2.9\times10^{-4}$ & 0.066 & 0.073 \\
 & 0.200--0.252 & 0.100--0.119 &  & 0.038 & 0.040 \\
 & 0.252--0.317 & 0.119--0.141 &  & 0.018 & 0.020 \\
 & 0.317--0.400 & 0.141--0.168 &  & $3.8\times10^{-3}$ & $9.6\times10^{-3}$ \\
 & 0.400--0.504 & 0.168--0.200 &  & $1.9\times10^{-4}$ & $4.1\times10^{-3}$ \\
 & 0.504--0.635 & 0.200--0.238 &  & $3.3\times10^{-6}$ & $1.7\times10^{-3}$ \\
 & 0.635--0.800 & 0.238--0.283 &  & $3.7\times10^{-8}$ & $6.6\times10^{-4}$ \\
 &  & 0.283--0.336 &  &  & $3.0\times10^{-4}$ \\
 &  & 0.336--0.400 &  &  & $2.0\times10^{-5}$ \\
 &  & 0.400--0.476 &  &  & $1.3\times10^{-6}$ \\
 &  & 0.476--0.566 &  &  & $1.5\times10^{-7}$ \\
 &  & 0.566--0.673 &  &  & $2.7\times10^{-8}$ \\
 &  & 0.673--0.800 &  &  & $4.2\times10^{-9}$ \\
\tableskip\tableline
\end{tabular}
\end{center}
NOTES.---%
We scale the $W_j$ by $(N_k/19)^{1/2}$ to remove the predicted scaling that
occurs when one refines the binning in wave number.
\end{table}

\begin{figure}[p]
\centerline{\plotone{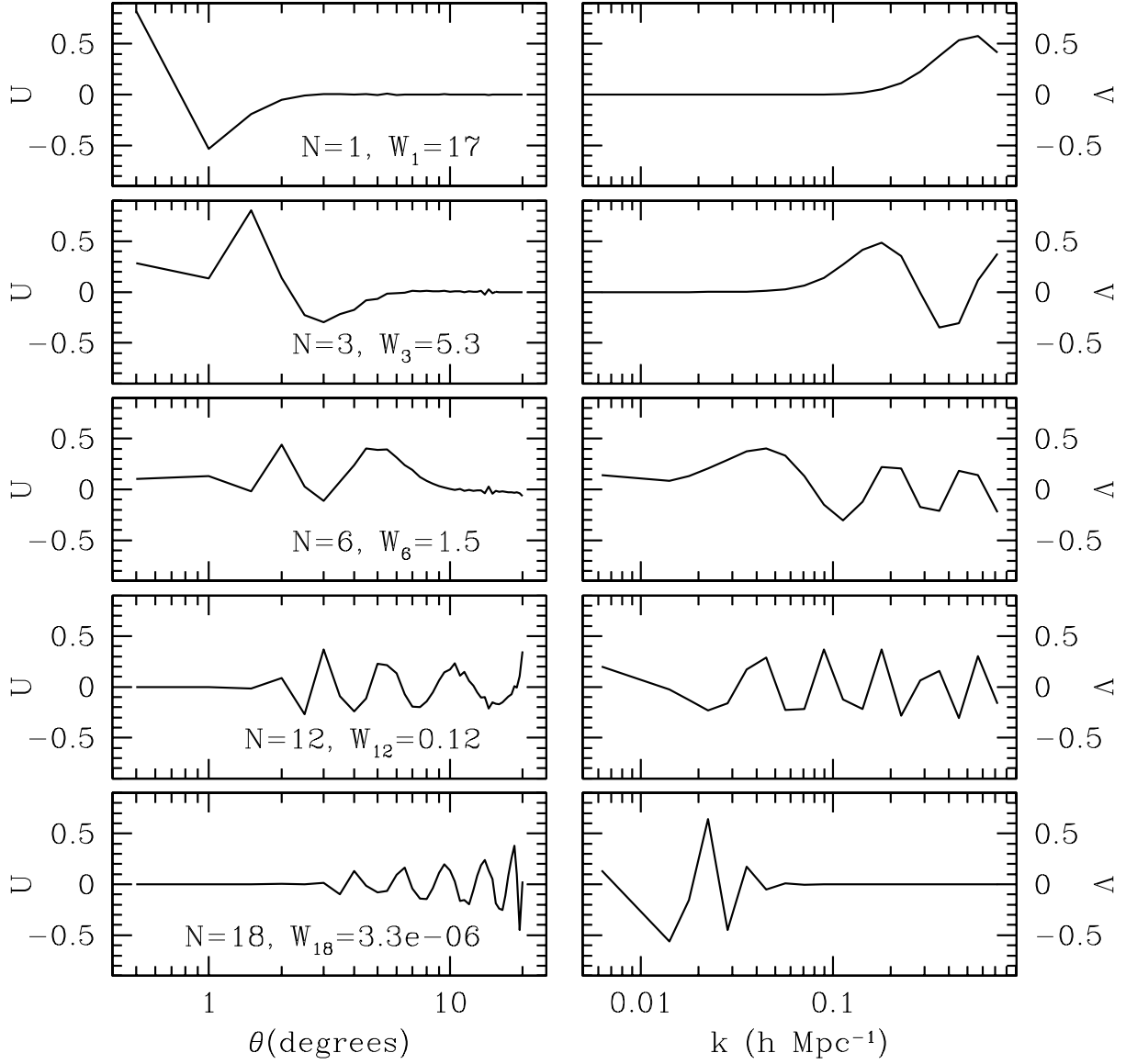}}
\caption{\label{fig:svuv}
Selected pairs of columns from the $U$ and $V$ matrices.  The $U$
column indicates the overlap with the $\bfw'\equiv C_w^{-1/2}\bfw$ 
data, while the $V$ column indicates the impact on $\bfP$ (as
normalized by $\bfPnorm$).  The singular value of each pair is also given.
The noise in the last $U$ vector is the result of the linear binning
in $\theta$.
}
\end{figure}

With $C_w$ and $\bfPnorm$, we can construct the kernel $G'$ and 
find its singular value decomposition.
Table \ref{tab:binning} shows the spectrum of singular values for
each of these 3 choices of binning.  
When one corrects by $N_k^{1/2}$
to account for the default scaling in $W_j$,
the large singular values are very stable as the binning is refined.  
Increasing the number of bins only increases the number of very small singular
values.  This demonstrates that 19 bins is a fine enough grid
to characterize the power spectrum; increasing to 25 bins would
only add degrees of freedom that are unconstrained by the angular data.
We could have put the factor of $N_k^{1/2}$ into the definition
of $\bfPnorm$ so as to make the large $W_j$ stable against changes
in binning, but since we will hereafter work only with $N_k=19$,
we opted against the extra complication.

\begin{table}[tb]\footnotesize
\caption{\label{tab:svuse}}
\begin{center}
{\sc Overlap of Singular Values with APM Data\\}
\begin{tabular}{cccc}
\tableskip\tableline\tableline\tableskip
$W_j$ & $U_j^T\bfw'$ & $W_j^{-1} U_j^T\bfw'$ & $W_{j,\rm eff}^{-1} U_j^T\bfw'$ \\
\tableskip\tableline\tableskip
17.1 & 21.2 & 1.2 & 1.2\\
9.3 & 7.8 & 0.84 & 0.84\\
5.3 & 9.0 & 1.7 & 1.7\\
3.2 & 4.4 & 1.4 & 1.4\\
2.0 & -4.1 & -2.0 & -2.0\\
1.5 & 2.3 & 1.6 & 1.6\\
1.1 & -1.1 & -1.0 & -1.0\\
0.88 & -0.56 & -0.64 & -0.64\\
0.59 & -1.5 & -2.6 & -2.6\\
0.37 & -0.79 & -2.1 & -1.6\\
0.22 & -0.96 & -4.4 & -1.9\\
0.12 & -1.5 & -12.4 & -3.1\\
0.066 & -0.30 & -4.5 & -0.60\\
0.038 & -0.56 & -14.9 & -1.1\\
0.018 & 0.082 & 4.5 & 0.16\\
$3.8\times10^{-3}$ & 1.2 & $3.3\times10^{2}$ & 2.5\\
$1.9\times10^{-4}$ & -0.31 & $-1.6\times10^{3}$ & -0.61\\
$3.3\times10^{-6}$ & 0.36 & $1.1\times10^{5}$ & 0.72\\
$3.7\times10^{-8}$ & -1.6 & $-4.2\times10^{7}$ & -3.1\\
\tableskip\tableline
\end{tabular}
\end{center}
NOTES.---%
The singular values are listed as $W_j$.
$U_j^T\bfw'$ shows the dot product between the
$j$th column of the $U$ matrix and the data vector $\bfw'$.
$W_{j,\rm eff}$ is the value of the $j$th SV rounded up to a minimum
value of $SV_C=0.5$.
\end{table}

As described in \S\ \ref{sec:svd}, the matching columns $U_j$
and $V_j$ map fluctuations in $\bfw'$ to those in $\bfP'$ with
an amplitude equal to the inverse of the singular value $W$.
Figure \ref{fig:svuv} displays 5 pairs of columns from the SVD.  
One sees that the large $W_j$ are associated with
small angular scales in $\bfw'$ and with smooth, large $k$ excursions
in $\bfP'$.  As the $W_j$ decrease, the oscillations become wilder
and move to larger angular scales.  The $U_j$ and $V_j$ vectors for
large $W_j$ remain very similar as one changes from $k19$ to $k25$
binning.

In Table \ref{tab:svuse}, we look at the overlap of these SV modes
with the APM $\bfw'$.  The quantity $U_j^T\bfw'$ is the number of
standard deviations by which the $j$th mode is demanded by $\bfw'$.
Dividing that by $W_j$ yields the amplitude of the effect on the
power spectrum (in units of $\bfPnorm$).  Modes with 
$U_j^T\bfw'\lesssim1$ are not statistically significant, while 
modes with $W_j^{-1}U_j^T\bfw'\gg1$ put enormous fluctuations 
in the power spectrum that are probably unphysical.  
One sees that the first 6 modes are clearly demanded, while the
remainder are of marginal significance.  We will include
8 modes in our quoted results, because this seems to be the 
transition between smooth and an oscillatory reconstruction.
However, we will also perform fits to CDM models with all modes
included in $\bfP'$.  In this regard, we also quote 
$W_{j,\rm eff}^{-1}U_j^T\bfw'$ in Table \ref{tab:svuse}, where
$W_{j,\rm eff}$ is $W_j$ rounded up to $SV_C=0.5$.  This is
to remind the reader that the value of $W_j$ used in $C_P$
is the one used in $\bfP'$.

\begin{figure}[p]
\centerline{\plotone{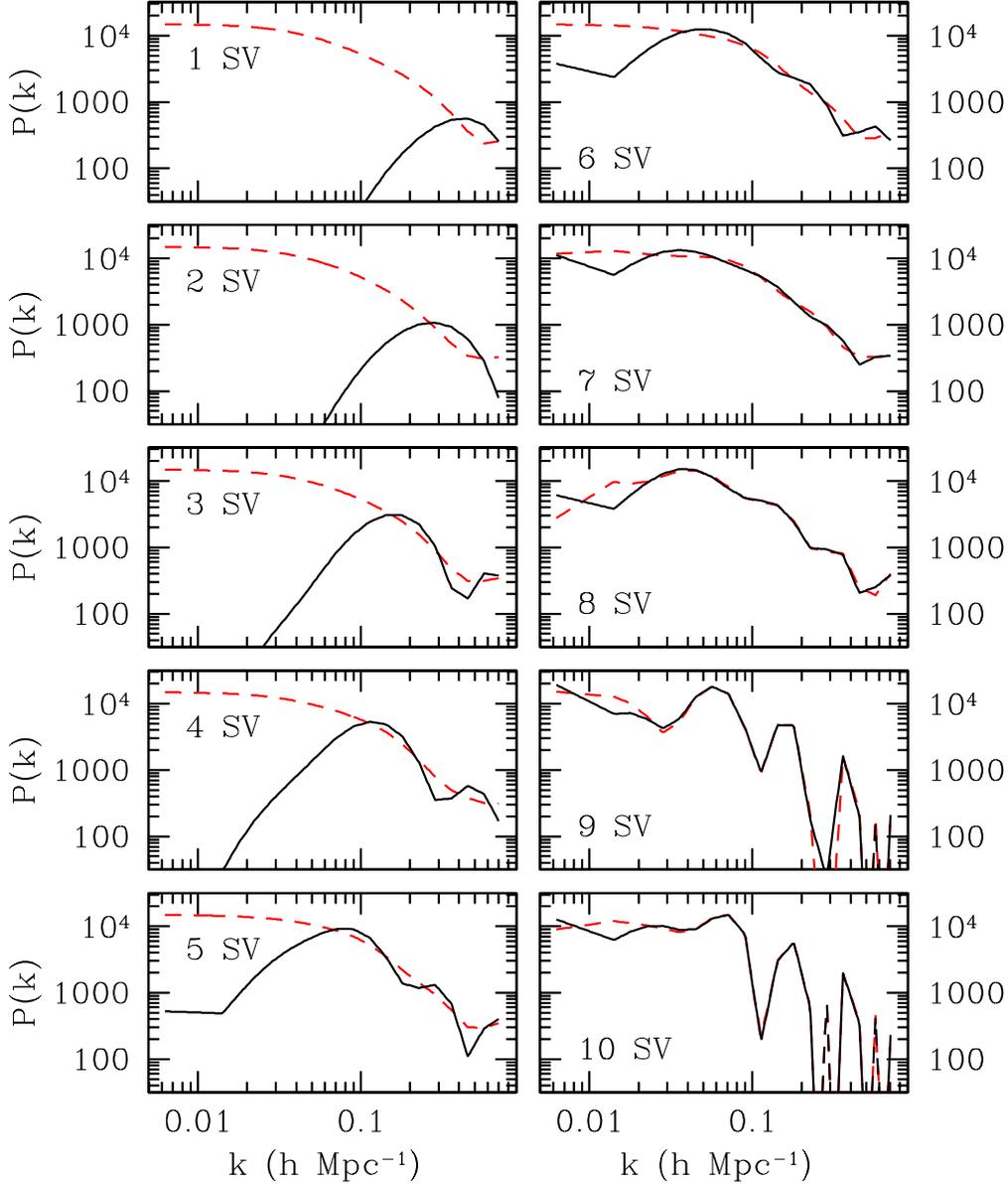}}
\caption{\label{fig:powsv}
The evolution of $P(k)$ as we include smaller singular values.
({\it solid}) The results with $\Pbias=0$.
({\it dashed}) The results with $\Pbias=P_{\rm norm}$.
The fact these two are identical with 8 SV shows that the resulting
power spectrum is not being biased by the SVD reconstruction.
}
\end{figure}

In Figure \ref{fig:powsv}, we show how the reconstructed power
spectrum develops as we add more SV modes.  The largest $W_j$ contribute
mostly small-scale power.  With 6 or 8 modes, a fairly smooth
shape appears that matches the expected form of $P(k)$.  
Adding smaller modes quickly 
makes the spectrum more oscillating, even with the artificial
increase in the $W_j$.  As Table \ref{tab:svuse}
shows, these oscillations are not statistically significant.

Figure \ref{fig:powsv} also shows the reconstructed power
spectrum if one chooses $\Pbias=P_{\rm norm}$.  Recall that 
$P_{\rm norm}$ was chosen to have a large amount of power on
large scales.  This means that any bias in $\bfP'$ due to the
alteration of SV will pull the spectrum toward high $P$ rather
than $P=0$.  This allows us to determine how many SV must be 
included to avoid bias in the large-scale power spectrum.
One sees that with 8 modes, the two power spectra are indistinguishable
at $k>0.02\ihmpc$.  This demonstrates that the smooth portion of
the power spectrum is being reconstructed in an unbiased way by
the SVD method.  In different words, the first 8 modes are all that
is needed to describe a smooth power spectrum like $P_{\rm norm}$.
Note that this does not mean that features in the resulting
power spectrum are statistically significant; that depends on the
covariance matrix $C_P$.

\begin{table}[tb]\footnotesize
\caption{\label{tab:power}}
\begin{center}
{\sc Reconstructed APM Power Spectrum \\}
\setlength{\tabcolsep}{3pt}
\begin{tabular}{cccc}
\tableskip\tableline\tableline\tableskip
$k$ Range & $P(k)$ & $(C_P)_{jj}^{1/2}$ & $(C_P^{-1})_{jj}^{-1/2}$ \\ 
\tableskip\tableline\tableskip
0.0000--0.0125 & 6088 & 22557 & 18091 \\
0.012--0.016 & 3802 & 27469 & 24706 \\
0.016--0.020 & 6127 & 26254 & 22338 \\
0.020--0.025 & 9354 & 24806 & 19804 \\
0.025--0.032 & 12891 & 22724 & 16836 \\
0.031--0.040 & 15175 & 19909 & 13607 \\
0.040--0.050 & 14625 & 17426 & 10874 \\
0.050--0.063 & 11458 & 14855 & 8324 \\
0.063--0.079 & 7724 & 11813 & 5643 \\
0.079--0.100 & 5544 & 9155 & 3474 \\
0.100--0.126 & 5077 & 6948 & 2061 \\
0.126--0.159 & 4331 & 5151 & 1210 \\
0.159--0.200 & 2394 & 3827 &  710 \\
0.200--0.252 &  978 & 2731 &  417 \\
0.252--0.317 &  936 & 2053 &  247 \\
0.317--0.400 &  776 & 1438 &  147 \\
0.400--0.504 &  206 & 1055 & 90.6 \\
0.504--0.635 &  252 &  855 & 61.0 \\
0.635--0.800 &  401 &  422 & 55.2 \\
\tableskip\tableline
\end{tabular}
\end{center}
NOTES.---%
This is the best-fit power spectrum using 8 SV to construct $\bfP'$ and
$SV_C=0.5$ in $C_w$.  The fit to the observed $P_2$ 
(eq.~\protect\ref{eq:P2fit}) was used to calculate $C_w$.
The units on wavenumber are
$\ihmpc$ and on power are $\hmpcC$.
Also shown are the diagonal elements of $C_P$ and $C_P^{-1}$, 
converted to give a standard deviation on $P(k)$.  These are
not very useful without the correlations in Table \protect\ref{tab:pcovinv}.
However, they do give the uncertainty on a single $k$ bin when 
marginalizing and not marginalizing, respectively, over all others.
In detail, this is the power at $z=0.11$ in an $\Omega_m=1$ cosmology.  
The power spectrum (and errors) would increase by about $20\%$
in a $\Lambda=0.7$ cosmology due to extra dilution of the
angular clustering caused by the additional volume at higher $z$.
The corrections in an open cosmology are $\sim\!3\%$.
\end{table}

\begin{table}[p]\footnotesize
\caption{\label{tab:pcovinv}}
\begin{center}
{\sc Reduced Covariance Matrix and Inverse Covariance Matrix for $P(k)$ \\}
\setlength{\tabcolsep}{3pt}
\begin{tabular}{ccccccccccccccccccc}
\tableskip\tableline\tableline\tableskip
1.00 & -0.24 & -0.27 & -0.25 & -0.15 & 0 & 0.10 & 0.08 & 0.01 & -0.05 & -0.02 & 0.03 & 0.01 & -0.02 & 0 & 0.01 & -0.01 & 0 & 0.01\\
  & 1.00 & -0.10 & -0.09 & -0.07 & -0.03 & 0.01 & 0.02 & 0.01 & -0.01 & -0.01 & 0 & 0 & 0 & 0 & 0 & 0 & 0 & 0\\
1.00 &   & 1.00 & -0.13 & -0.11 & -0.07 & -0.02 & 0.01 & 0.02 & 0 & -0.01 & 0 & 0.01 & 0 & 0 & 0 & 0 & 0 & 0\\
0.39 & 1.00 &   & 1.00 & -0.16 & -0.14 & -0.08 & -0.02 & 0.03 & 0.02 & -0.01 & -0.02 & 0.01 & 0.01 & -0.01 & 0 & 0 & -0.01 & 0\\
0.44 & 0.29 & 1.00 &   & 1.00 & -0.24 & -0.19 & -0.08 & 0.03 & 0.06 & 0 & -0.03 & 0.01 & 0.02 & -0.01 & 0 & 0.01 & 0 & 0\\
0.43 & 0.30 & 0.38 & 1.00 &   & 1.00 & -0.32 & -0.19 & -0.02 & 0.07 & 0.03 & -0.04 & -0.01 & 0.03 & 0 & -0.02 & 0.01 & 0.01 & -0.01\\
0.33 & 0.26 & 0.35 & 0.46 & 1.00 &   & 1.00 & -0.32 & -0.16 & 0.01 & 0.07 & 0 & -0.04 & 0.01 & 0.02 & -0.02 & 0 & 0.02 & -0.02\\
0.18 & 0.17 & 0.27 & 0.41 & 0.56 & 1.00 &   & 1.00 & -0.37 & -0.16 & 0.06 & 0.08 & -0.04 & -0.03 & 0.03 & 0 & -0.02 & 0.01 & -0.01\\
0.05 & 0.09 & 0.18 & 0.33 & 0.50 & 0.65 & 1.00 &   & 1.00 & -0.45 & -0.13 & 0.11 & 0.06 & -0.07 & -0.02 & 0.05 & -0.01 & -0.03 & 0.04\\
-0.01 & 0.04 & 0.10 & 0.21 & 0.36 & 0.53 & 0.68 & 1.00 &   & 1.00 & -0.51 & -0.08 & 0.16 & 0.01 & -0.08 & 0.05 & 0.02 & -0.04 & 0.04\\
-0.02 & 0.01 & 0.04 & 0.09 & 0.19 & 0.32 & 0.50 & 0.72 & 1.00 &   & 1.00 & -0.56 & -0.04 & 0.17 & -0.02 & -0.08 & 0.04 & 0.03 & -0.05\\
-0.01 & 0 & 0.01 & 0.03 & 0.07 & 0.14 & 0.29 & 0.53 & 0.79 & 1.00 &   & 1.00 & -0.62 & 0.04 & 0.18 & -0.11 & -0.03 & 0.08 & -0.08\\
0 & 0 & 0 & 0.01 & 0.02 & 0.05 & 0.12 & 0.30 & 0.57 & 0.83 & 1.00 &   & 1.00 & -0.65 & 0.06 & 0.18 & -0.08 & -0.06 & 0.11\\
0 & 0 & 0 & 0 & 0.01 & 0.02 & 0.04 & 0.13 & 0.32 & 0.60 & 0.86 & 1.00 &   & 1.00 & -0.68 & 0.20 & 0.11 & -0.14 & 0.11\\
0 & 0 & 0 & 0 & 0 & 0.01 & 0.02 & 0.05 & 0.15 & 0.34 & 0.61 & 0.87 & 1.00 &   & 1.00 & -0.76 & 0.17 & 0.19 & -0.27\\
0 & 0 & 0 & 0 & 0 & 0 & 0.01 & 0.02 & 0.06 & 0.15 & 0.34 & 0.61 & 0.88 & 1.00 &   & 1.00 & -0.66 & 0.17 & 0.04\\
0 & 0 & 0 & 0 & 0 & 0 & 0 & 0.01 & 0.02 & 0.06 & 0.16 & 0.34 & 0.61 & 0.88 & 1.00 &   & 1.00 & -0.81 & 0.62\\
0 & 0 & 0 & 0 & 0 & 0 & 0 & 0 & 0.01 & 0.02 & 0.07 & 0.16 & 0.34 & 0.62 & 0.88 & 1.00 &   & 1.00 & -0.95\\
0 & 0 & 0 & 0 & 0 & 0 & 0 & 0 & 0 & 0.01 & 0.02 & 0.07 & 0.16 & 0.35 & 0.62 & 0.89 & 1.00 &   & 1.00\\
0 & 0 & 0 & 0 & 0 & 0 & 0 & 0 & 0 & 0 & 0 & 0.02 & 0.06 & 0.17 & 0.36 & 0.64 & 0.90 & 1.00 &  \\
0 & 0 & 0 & 0 & 0 & 0 & 0 & 0 & 0 & -0.01 & -0.01 & -0.01 & 0.01 & 0.05 & 0.17 & 0.41 & 0.72 & 0.94 & 1.00\\
\tableskip\tableline
\end{tabular}
\end{center}
NOTES.---%
The upper triangle shows $C_P$ after we have divided each row and column
by the square root of its respective diagonal element.
The lower triangle shows $C_P^{-1}$ with its diagonal divided out.
Both matrices are symmetric of course.
The square root of the diagonal of $C_P$ is given as the third column
of Table \ref{tab:power}; the inverse of the square root of the diagonal
of $C_P^{-1}$ is the last column in that chart.
Please note that well-constrained directions have small eigenvalues in
$C_P$ and large eigenvalues in $C_P^{-1}$.  With only two significant
figures, the small eigenvalues will be inaccurate.  
$C_P$ is quoted here only to show the correlation coefficients.
If one wishes to fit smooth power spectrum models using the above matrices,
one must use $C_P^{-1}$.  
To calculate the change in $\chi^2$ of a particular deviation in $P(k)$, 
divide each element of the vector of $\Delta P$'s 
by the corresponding number in the last
column of Table \ref{tab:power} and then contract a symmetrized version 
of the lower triangle of the matrix above 
by the vector of fractional variations (i.e. $v^T M v$).
Note that the sense of the off-diagonal terms of $C_P^{-1}$ is to
penalize non-oscillatory deviations in $P(k)$ more than the sum of
the significance in each $k$ would suggest.  Conversely, oscillatory
fluctuations in power are more permitted than the sum of significances
would suggest.  We use $SV_C=0.5$ here.
Contact the authors if more significant figures are needed.
\end{table}

We present the best-fit $P(k)$ using the largest 8 SV 
in Table \ref{tab:power}.
The reduced forms of $C_P$ and $C_P^{-1}$, using $SV_C=0.5$,
are shown in Table \ref{tab:pcovinv}, with the diagonal elements
in Table \ref{tab:power}.  
The reduced form of $C_P$ shows the correlation coefficients between
the $k$ bins.  Neighboring bins are anti-correlated with 
correlation coefficients ranging from -0.95 on small scales to
-0.5 on moderate scales to -0.1 on the largest scales.
While $C_P$ could be used to calculate the change in $\chi^2$
for particular excursions around $\bfP$, two significant figures
is not enough to do so correctly.  Instead, one should use the
quoted $C_P^{-1}$.  For smooth excursions around the best-fit $P(k)$, great
accuracy in $C_P^{-1}$ is not required.  However,
because the oscillatory excursions are more singular,
one should contact the authors for more significant figures if one
wishes to manipulate these kinds of fluctuations.

The $w(\theta)$ using this best-fit power spectrum differs from
the input $w(\theta)$ by $\chi^2=46$.  There are 40 bins in $\theta$
and 19 bins in $k$, so the naive number of degrees of freedom is 21.
However, we have included only 8 SV modes in constructing $P(k)$, so
in this sense there are 32 degrees of freedom.  
If we include all 19 modes, $\chi^2$ drops to 41.  
This decrease is small
because most of these modes have had their $W_j$ adjusted before use
in $\bfP'$.  This causes the modes to be smaller in amplitude than $C_w$
would suggest.  As we reduce $SV_C$, $\chi^2$ slowly drops as additional
modes reach their ``natural'' scale.

With 32 degrees of freedom, a $\chi^2$ of 46 is 5\% likely, and
hence the fit is only marginal.  This may indicate that our errors
are underestimated.  However, we find that changing
the power on small scales makes a large difference to $\chi^2$,
but almost no difference to the reconstruction on large scales
or the fits to CDM models.  For example, if we calculate $C_w$ using
the 2-dimensional projection of a $\Gamma=0.25$ CDM model with $\sigma_8=0.89$
and the non-linear corrections of \citet{Pea96}, $\chi^2$ drops to 22.  
Removing the non-linear corrections increases $\chi^2$ to 131.  
These two models bracket the observed $P_2$ on small scales.
Neither change to $C_w$ affects large-scale model fits at all.
We therefore conclude that it is our small-scale errors, 
not our large-scale ones, that are slightly underestimated.

\subsection{Constraints on $P(k)$ at large scales}\label{sec:cdm}

Having reconstructed the power spectrum and its covariance matrix,
we wish to consider how the results constrain the large-scale power
spectrum.  Large scales are important because the spectral 
signatures that would identify particular cosmologies are strongest
there.  Moreover, non-linear evolution erases any residual features
on small scales, at which point the potential problem of 
scale-dependent bias might further obscure the link to cosmology.
While the small-scale power spectrum is certainly important,
it is the large-scale power that can be most cleanly linked to cosmological 
parameters.

We begin by discussing two of the important phenomenological
results that have been associated with the APM power spectrum.
First, does the power reach a maximum at $k\approx0.04\ihmpc$ and 
drop at the larger scales \citep{Gaz98}?
One can see from the comparison of the two curves in Figure
\ref{fig:powsv} that any downturn in the power spectrum at 
$k<0.04\ihmpc$ is only contributed by modes 7 and higher.  With the
first 6 modes, the situation at $k<0.04\ihmpc$ is completely
prior-dominated.  Unfortunately, modes 7 and 8 have $U_j^T\bfw'=-1.1$ 
and -0.56 (Table \ref{tab:svuse}), respectively, and so they improve 
the $\chi^2$ of the
fit to the $w(\theta)$ data by only 1.52.  Modes 9 and higher produce
oscillations in $\bfP$ that are inconsistent with small-scale data.
Hence, we conclude that this suggestion of a downturn 
in $P(k)$ at $k<0.04\ihmpc$ is not statistically significant.

Another way to quote this significance is to look at how well the
covariance matrix constrains a constant power fluctuation in the
first 6 $k$ bins ($k<0.04\ihmpc$).  Contracting this submatrix
of $C_P^{-1}$ with the vector of 6 ones gives $3.4\times10^{-8}$,
which means that the 1-$\sigma$ limit on such an excursion is
$4500\hmpcC$.  
Using a submatrix of $C_P^{-1}$ corresponds to assuming perfect
information about smaller scales; in other words, this fluctuation
leaves the best-fit power at $k>0.04\ihmpc$ unchanged.  Allowing
the smaller scales to vary within their errors increases $\sigma$
to $5200\hmpcC$.
Using a $\Gamma=0.25$, $\sigma_8=0.9$ CDM model for $P_2$ only 
increases these errors.
Comparing to the best-fit $\bfP$, the hypothesis that $P(k)=15000\hmpcC$ 
on all scales below $k=0.04\ihmpc$
can only be rejected at 1.25- or 1.21-$\sigma$, using the assumptions of
perfect and imperfect small-scale information, respectively.
The 2-$\sigma$ upper bound on the power at $k<0.04\ihmpc$ is roughly $2\times10^4\hmpcC$.
Again we conclude that the downturn at $k<0.04\ihmpc$ is not significant.

The shape of the APM power spectrum at $k\approx0.1\ihmpc$ scales 
has been noted for a sharp break that does not fit simple CDM models
\citep{Gaz98,Gaw98}.  
Using the covariance matrix in Table \ref{tab:pcovinv},
we find that the BE93 power spectrum 
and a $\Gamma=0.25$, $\sigma_8=0.89$ CDM model 
differ by only 1.2-$\sigma$ at $k<0.2\ihmpc$ even if smaller scales are held fixed.
Alternatively, Table \ref{tab:power} shows that 20\% fluctuations in
power at $k\approx0.1\ihmpc$ are permitted.  We therefore find that
the shape of the BE93 power spectrum at these scales 
is not statistically different from that of the CDM model.

Unfortunately, the large anti-correlated errors in the spatial
power spectrum makes it difficult to visualize the constraints 
at large scales.  We therefore fit a set of theoretical power spectra 
to the power spectrum and study the resulting constraints on the
parameter space.
For this we consider a very
restricted set, namely a scale-invariant CDM model specified by
$\Gamma$ and an amplitude $\sigma_8$.  We include non-linear evolution
according to the formulae of Peacock \& Dodds (1996), but one
should note that this means that we have assumed that the galaxies
are unbiased with respect to the mass. 
We include only wavenumbers less than $k<k_c$.  We use $k_c=0.2\ihmpc$
in most cases.  This is roughly the transition point between the
linear and non-linear regimes, which is where the problems of
scale-dependent bias could appear and where our Gaussian assumption
in computing the sample variance will begin to be overly optimistic.
We have marginalized over the smaller scales when computing $\chi^2$;
however, we get similar constraints on large scales if we
hold the small-scale power spectrum equal to a power law $P\propto k^{-1.3}$
with unknown amplitude.

\begin{figure}[p]
\centerline{\plotone{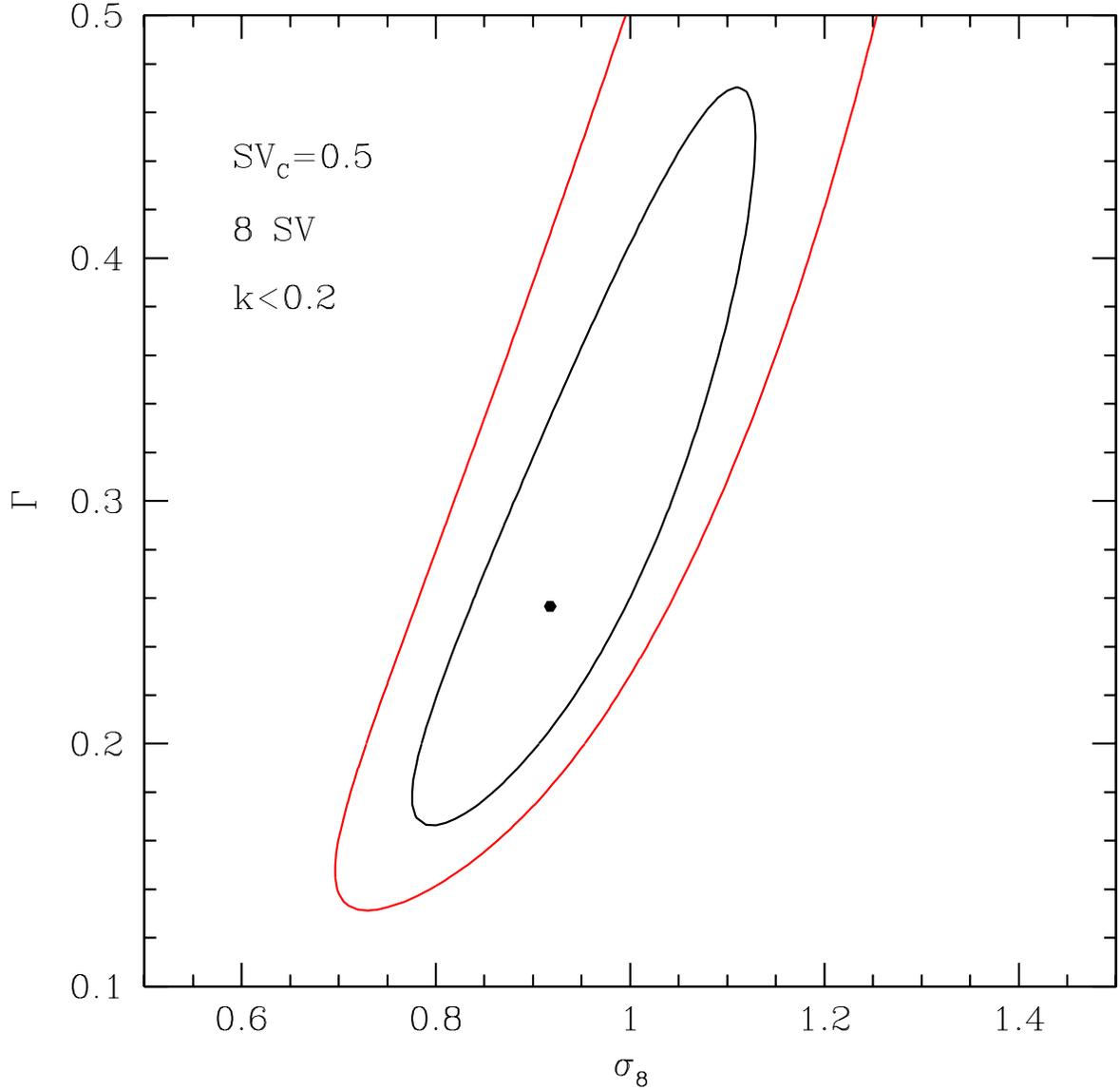}}
\caption{\label{fig:pc1_1}
Constraints on a 2-parameter family of CDM power spectra when fit
to the best-fit power spectrum of Table \protect\ref{tab:power}.  
Non-linear theoretical power spectra are used \protect\citep{Pea96}.
$\bfP'$ and $C_w$ are calculated
using the observed $P_2$ values (eq.~\protect\ref{eq:P2fit}).
$W_j$ less than $SV_C=0.5$ have been increased to 0.5 in constructing
$C_w$, and only the first 8 SV modes have been included in $\bfP'$.
All SV modes are used in $C_w$.
The difference between this reconstruction
and the model is then used to find $\chi^2$.  
68\% and 95\% contours ($\Delta\chi^2=2.30$ and 5.41) are shown.  
Only wavenumbers $k<0.2\ihmpc$ are used; 
we marginalize over the uncertainty at larger wavenumbers.
}
\end{figure}

\begin{figure}[p]
\centerline{\plotone{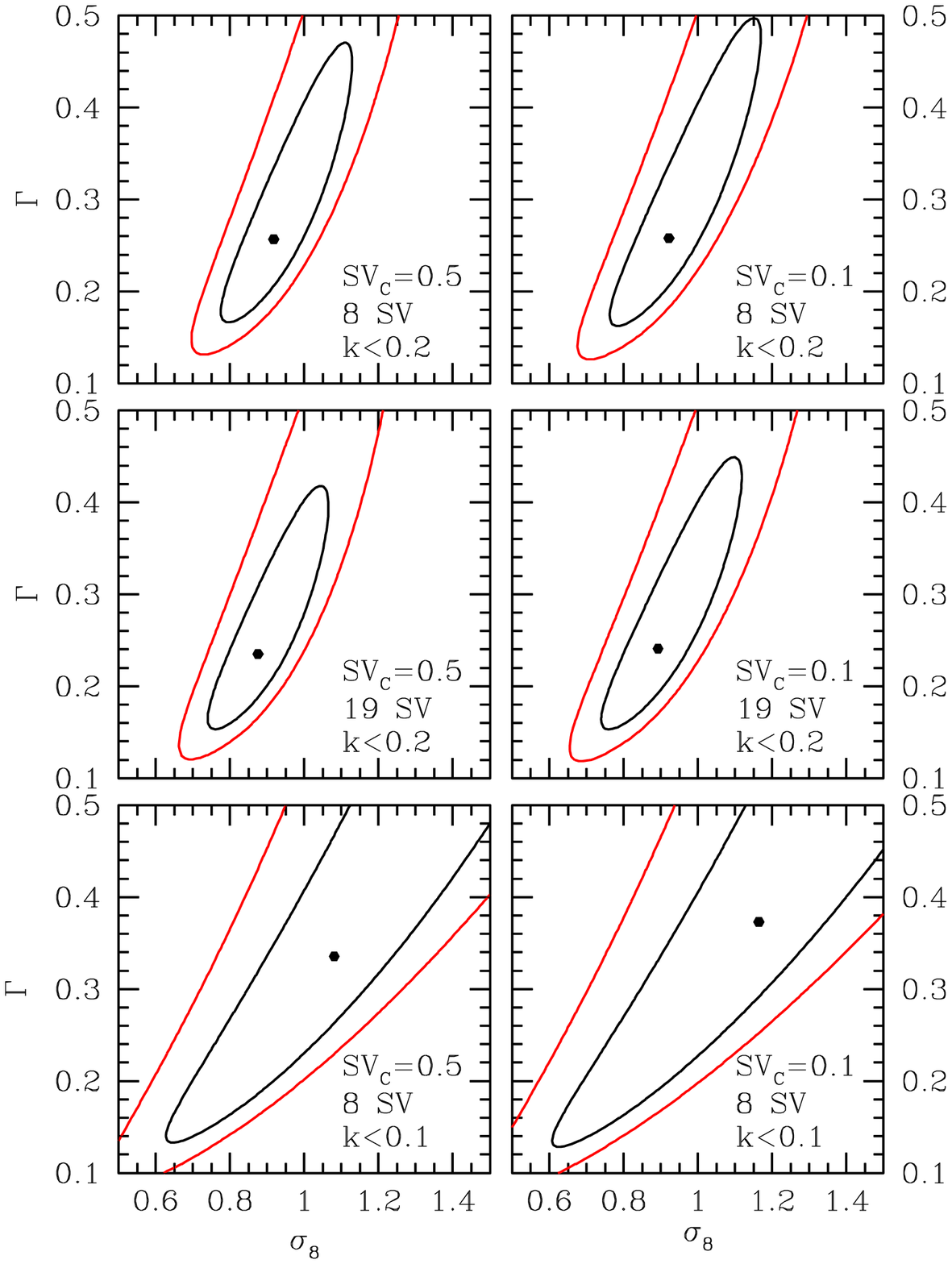}}
\caption{\label{fig:pc6_5}
As Figure \protect\ref{fig:pc1_1} but with changes to parameters
of the reconstruction and fit.
({\it left panels}) $W_j$ less than $SV_C=0.5$ are increased to 0.5.  
({\it right panels}) $W_j$ less than $SV_C=0.1$ are increased to 0.1.  
({\it top panels}) Only the 8 modes with the largest $W_j$ are used 
in constructing the
power spectrum.  All modes are used in 
constructing the covariance matrix.
Only wavenumbers $k<0.2\ihmpc$ are used; 
we marginalize over the uncertainty at larger wavenumbers.
({\it middle panels}) As top, but all 19 SV are used 
in constructing the power spectrum.
({\it bottom panels}) As top, but only $k<0.1\ihmpc$ is used.
}
\end{figure}

In Figure \ref{fig:pc1_1}, we show the constraints on these CDM models.
We use $SV_C=0.5$ and include 8 modes in calculating $\bfP$.  All modes
are used in calculating $C_P$.  The contours are
drawn at $\Delta\chi^2=2.30$ and 5.41, which are the values for a 68\%
and 95\% confidence region in a 
Gaussian ellipse.\footnote{In detail, the actual integral of the 
probability would differ from
this, but we neglect this effect because it wouldn't alter the
basic point and would make the results depend on one's choice of
metric in parameter space.}
The constraints are rather loose.  The most likely model has
$\Gamma=0.26$ and $\sigma_8=0.92$.  If we marginalize over
$\sigma_8$, $\Gamma$ has a range of 0.19--0.37 (68\%) and
0.15--0.58 (95\%).  
The strong skewness of the constraint region towards higher $\Gamma$ is 
an artifact of using $\Gamma$ as a parameter.
Increasing $\Gamma$ removes large-scale power, but since the error
bars are not changing with $\Gamma$, eventually a small change in 
power maps to a large change in $\Gamma$.
Because of this skew tendency, we are not generally concerned about
modest changes in the upper limit on $\Gamma$ range, as they correspond
to small changes in the actual power.

If we use all 19 SV modes in constructing $\bfP$, the $\chi^2$
for the best-fit CDM model is 6.  This is based on 11 degrees
of freedom, as calculated from 13 $k$ bins and 2 parameters.
Alternatively, one can think of this as 19 $k$ bins and 8 parameters:
the 2 CDM parameters at $k<0.2\ihmpc$ and 6 bins of bandpower
at $k>0.2\ihmpc$ that have been marginalized over.  $\chi^2=6$ on
11 degrees of freedom is small but not statistically abnormal.
We would therefore say that the CDM model is an acceptable fit
to the data.

If one uses only 8 SV modes to construct $\bfP$, the best-fit
CDM model has $\chi^2=0.5$.  One might imagine that with 8 modes
and 8 parameters, one has zero degrees of freedom.  However, the
other 11 modes haven't been removed from the $\chi^2$; they have 
simply had their amplitude in $\bfP$ set to zero.  If all of the models had
zero overlap with the omitted modes, then we would indeed lose
one degree of freedom per frozen mode.  However, the overlap is
small---because the omitted modes are wiggly while the models are
smooth---but non-zero.  Hence, we do not find the small $\chi^2$ to
be surprising, but it is difficult to say this quantitatively.

One might worry that allowing the power at $k>0.2\ihmpc$ to vary within its
errors could cause great uncertainty on large scales because we haven't
included any angular data on scales below $0^\circ.5$.  One way to 
address this is to force the small-scale power spectrum to a smooth
form.  Holding the power at $k>0.36\ihmpc$ equal to a $k^{-1.3}$ power law
of unknown amplitude has only a small effect on the allowed region for $\Gamma$.
Extending this power-law to $k=0.2\ihmpc$ causes the confidence
intervals on $\Gamma$ to be 0.19--0.33 (68\%) and 0.155--0.50 (95\%).  
This is a minor improvement for such a strong prior.  
As second test, we attempt to include our knowledge of the small-scale 
power spectrum directly in the inversion by
replacing the DG99 $w(\theta)$ at $\theta<2^\circ)$ with a finely-sampled 
representation of the BE93 fitting form to $w(\theta)$ that extends to
$0^\circ.07$.  This yields confidence regions on $\Gamma$ of 0.185-0.36 (68\%) 
and 0.145-0.57 (95\%).
Hence, we conclude that the small scales are well-enough constrained by
angular data at $\theta>0^\circ.5$ that their uncertainties do not
affect the reconstruction of the power spectrum at $k<0.2\ihmpc$.

In Figure \ref{fig:pc6_5}, we vary some of the above assumptions.
The top row of the Figure shows the results as we
vary $SV_C$.  The left panel is $SV_C=0.5$, as in Figure \ref{fig:pc1_1}.
In the right panel, we use $SV_C=0.1$.  This gives the ill-constrained
directions in the power spectrum fit 5 times more freedom.  Indeed,
their amplitudes will commonly exceed unity, which is unphysical for a
positive-definite quantity like the power spectrum.  The constraints on
CDM parameters are slightly worse, but not considerably so.  
Reducing $SV_C$ even more makes little difference.  Increasing 
$SV_C$ above 1.0 begins to shrink the allowed region and move
the best-fit point to higher $\Gamma$.  This is because the
modes with $W_j\gtrsim2$ contribute little large-scale power;
if all the modes with smaller SV are suppressed by setting $W_j=SV_C$,
then the result becomes biased toward zero power on large scales.

The middle row of Figure \ref{fig:pc6_5} shows the results when
all SV are included in calculating the best-fit power
spectrum.  Generally the differences are small, showing that these 
smaller SV have little effect on fits to CDM models.
Larger $\Gamma$ are slightly less favored, but the constraints
are still very broad.  One should remember that since the small
$W_j$ have been increased to $SV_C$ before being added to $\bfP$,
adding such modes is not more ``correct'' in the sense of
yielding an unbiased estimator or returning the best-fit 
(and non-positive) power spectrum.

The bottom row of Figure \ref{fig:pc6_5} restricts the fit to 
even larger scales, $k_c=0.1\ihmpc$.  The constraints are
considerably worse: in particular, no interesting upper bound
can be set on $\Gamma$.  The best-fit $\Gamma$ is also higher.

With $k_c=0.2\ihmpc$, the tilt of the constraint region in the 
$\Gamma$--$\sigma_8$ plane is in the sense of a tight constraint
on the \rms\ fluctuations on a larger scale.  That is, if we were
to plot the constraint on the $\Gamma$--$\sigma_{24}$ plane, the
region would be roughly perpendicular to the axes.
This is not surprising, because $\sigma_8$ is dominated by 
$k\approx0.2\ihmpc$, and the fit should focus on larger
scales.  

Our fit to the observed $P_2$ approaches a constant as $K\rightarrow0$,
which means that it does not approach scale-invariance ($P_2\propto K$)
on the largest scales.  One might worry that this causes an overestimate
of the errors on large scales.  We can address this by using a CDM 
power spectrum when calculating $C_w$.  We take a model
with $\Gamma=0.25$ and $\sigma_8=0.89$ and  
project the non-linear $P_3$ to $P_2$.
While the CDM $P_2$ does eventually go to zero at large scales, it
actually exceeds our fit to observations at $K=10$.
Using the CDM $P_2$ to calculate $C_w$, we find constraints in the
$\Gamma$--$\sigma_8$ plane that are a very close match to those in
Figure \ref{fig:pc6_5}.  In detail, the best-fit $\Gamma$ and the 
confidence intervals shift by only 0.01, which is far within the errors.

When fitting to cosmological models, one can include the fact that 
the sample-variance portion of the covariance matrix depends on the
model itself.  For example, one might worry that large $\Gamma$ models
would predict smaller sample variance and hence be less favored than
Figure \ref{fig:pc1_1} would suggest.  We therefore repeat our fits 
to CDM models, using the
model at each point to generate the covariance matrix and the best-fit
power spectrum.  We then calculate $\chi^2$ as before.
We find that the confidence regions are essentially unchanged and
that the best-fit $\Gamma$ moves by less that 0.01.

\subsection{Higher-order terms in the Covariance} 
\label{sec:nongauss}

In our treatment so far,
we have only included the Gaussian terms in the
covariance matrix $C_w(\theta,\theta')$. 
We want to estimate the size of the non-Gaussian terms and
determine if their inclusion could substantially change our results. To
do this, we will use the hierarchical ansatz for the higher-order
moments of the density field.  The four-point function is assumed to be
\begin{eqnarray}\label{t41}
T_4({\Kvect}_1,{\Kvect}_2,{\Kvect}_3,{\Kvect}_4)&=&  r_a\left[P_2({\vec
k}_1)P_2({\Kvect}_2)P_2({\Kvect}_{13})+ {\rm cyc.}\right] + \nonumber 
\\&& r_b \left[P_2({\vec
k}_1)P_2({\Kvect}_2)P_2({\Kvect}_{3})+ {\rm cyc.}\right], 
\end{eqnarray}
where $r_a$ and $r_b$ are constants describing the hierarchical 
amplitudes for the two
different topologies of diagrams contributing to the four-point
function.
With the same set of approximations we used to obtain equation
(\ref{eq:Cww}), the full covariance of $w(\theta)$ is
\beqa\label{eq:Cwwng}
C_w(\theta,\theta') &=&
{1\over A_\Omega} \left[\int_0^\infty {dK\,K \over 2\pi}\ 2\ 
 P_2^2(K) J_0(K\theta)
J_0(K\theta') \nonumber + \right.\\ && 
\left.\int_0^\infty {dK\,K \over 2\pi} 
\int_0^\infty {dK'\,K'\over 2\pi} \bar T_4(K,K') J_0(K\theta)
J_0(K'\theta')\right], \nonumber \\
\nonumber \\
\bar T_4(K,K')&=&\int {d^2K_a \over A_{r}} \int {d^2K_b \over A'_{r}}
T_4({\Kvect}_a,-{\Kvect}_a,{\Kvect}_b,-{\Kvect}_b).
\eeqa
The last integral is an angular average of the four-point function
over rings in $K$ space of area $A_r$ centered around $K$ and $K'$.

\begin{figure}[p]
\centerline{\plotone{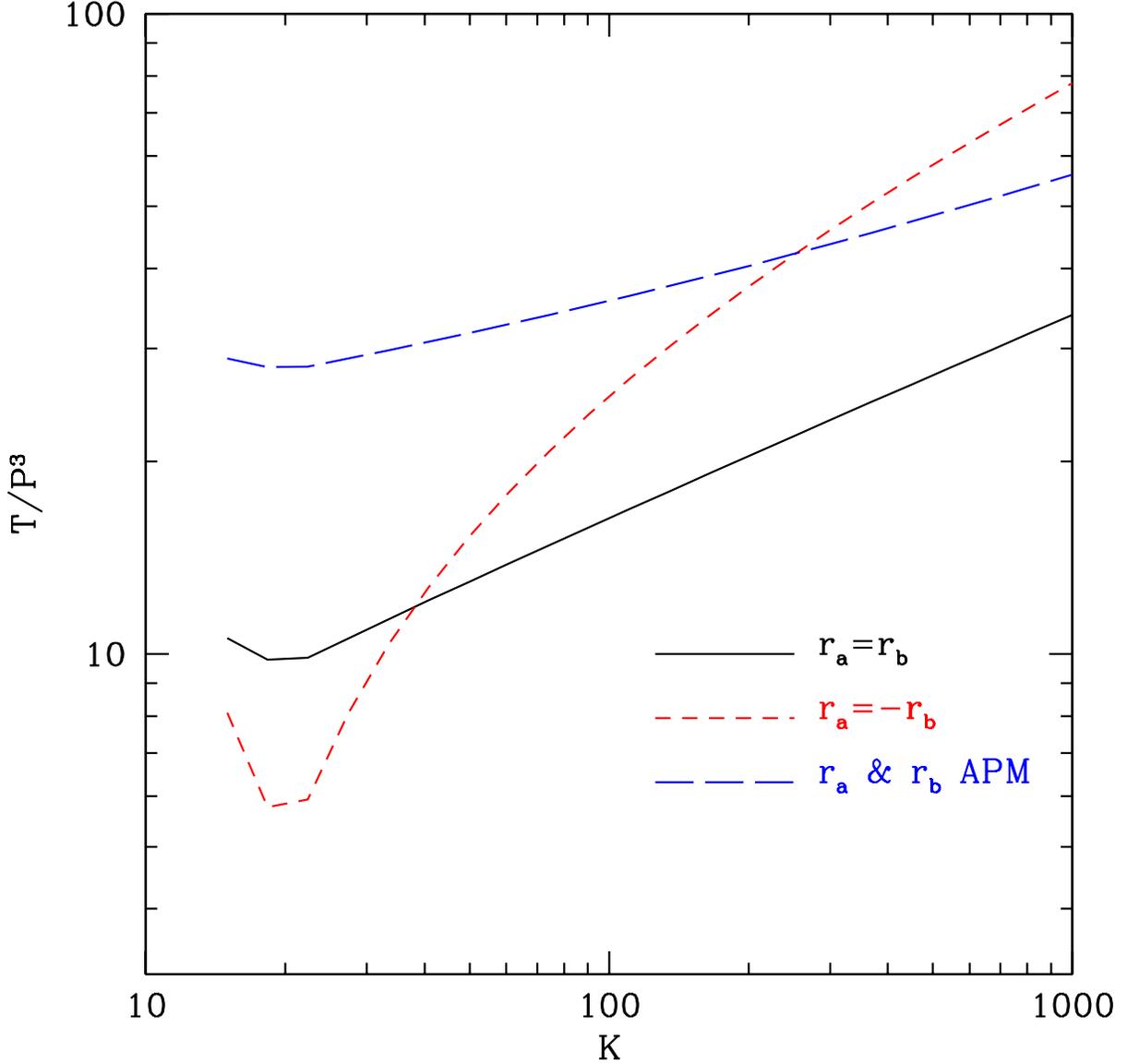}}
\caption{\label{tp3ratio}
$T(K,K)/P_2^3(K)$ in the hierarchical model for different
choices of $r_a$ and $r_b$. The hierarchical amplitudes measured from
APM \protect\citep{SzSz97} are expected to be larger than 
the amplitudes relevant for the
configurations that determine the variance of the power spectrum. The
curves for $r_a=\pm r_b$ are each normalized to the $T/P^3$ values
($R=4[2r_a+r_b]=12$) calculated when the spatial quantities obtained in 
$N$-body simulations were projected to the angular quantities using the
APM selection function.}
\end{figure}

In order to estimate the size of this contributions we need to have an
estimate of the hierarchical amplitudes $r_a$ and $r_b$. \cite{SzSz97}
estimated these amplitudes for APM by measuring 
two different configurations of the four-point function. 
They obtained $r_a=1.15$ and $r_b=5.3$. The diagonal terms 
$\bar T_4(K,K)$
are determined mainly by the
combination $R=4(2r_a+r_b)=30.4$. 
In \citet{Sco99}, it was shown that the hierarchical ansatz is not
a particularly good approximation for the configurations of the four-point 
function relevant for the variance of the power spectrum (or the
two-point function), i.e.\ those configurations in which two pairs of $\Kvect$
add up to zero.  The amplitudes of the important
configurations were roughly a factor of five smaller than one would
naively expect. Therefore a smaller value of the hierarchical
coefficients should be used for the variance calculation. The spatial 
statistics 
measured in particle-mesh $N$-body simulations imply after projection 
that the hierarchical coefficients for APM should satisfy
$4(2r_a+r_b)\approx 12$.
Figure \ref{tp3ratio} shows the ratio of $T/P^3$ along the diagonal
for the different choices of $r_a$ and $r_b$. 

\begin{figure}[p]
\centerline{\plotone{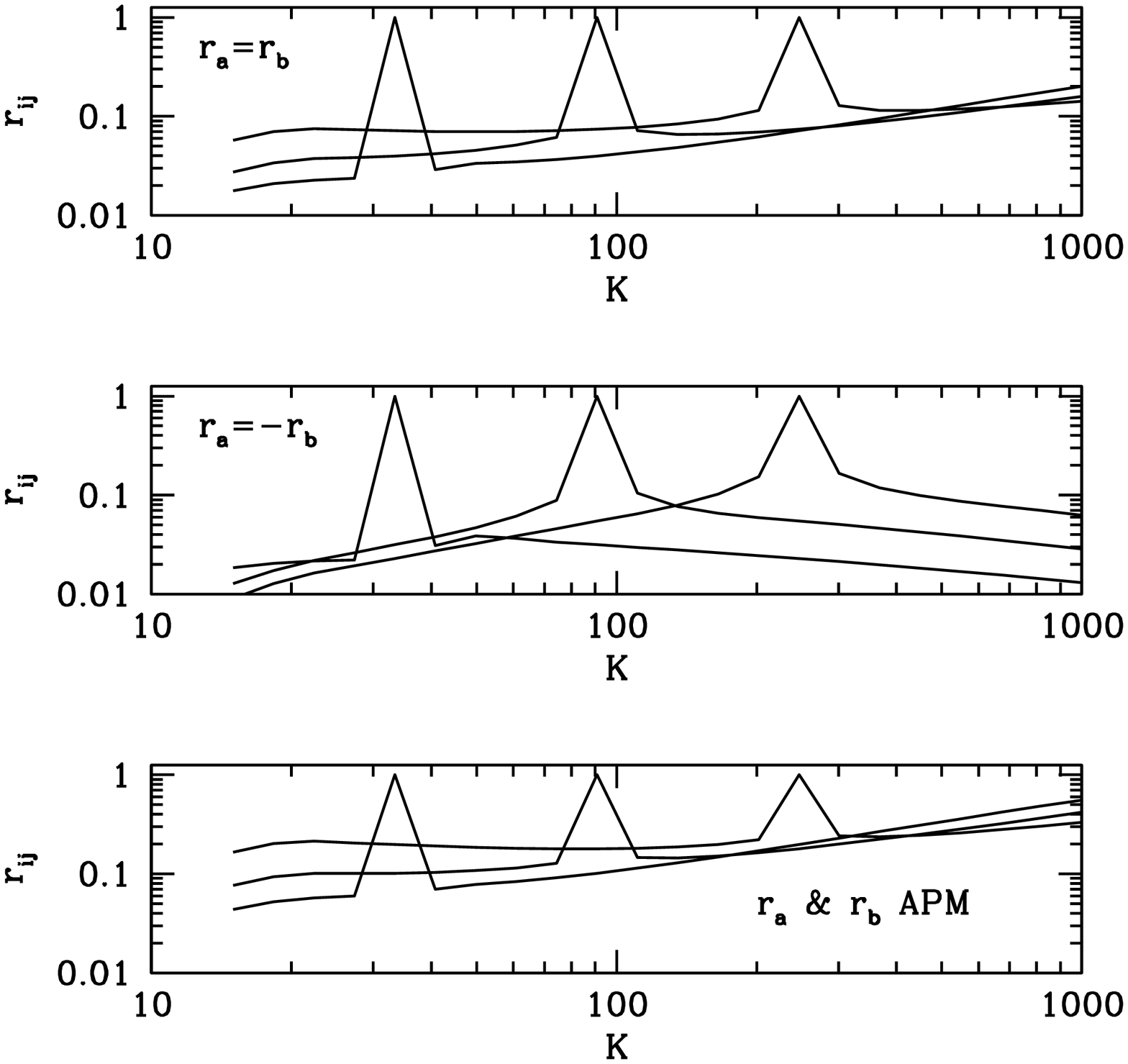}}
\caption{\label{rij}
Cross-correlation coefficients between different $K$ shells of 
the angular power spectra for the choices of $r_a$ and $r_b$
listed in Figure \protect\ref{tp3ratio}. 
Each curve shows the 
cross-correlation coefficient between one $K$ shell (the one with $r_{ij}=1$) 
and all the rest. }
\end{figure}

\begin{figure}[p]
\centerline{\plotone{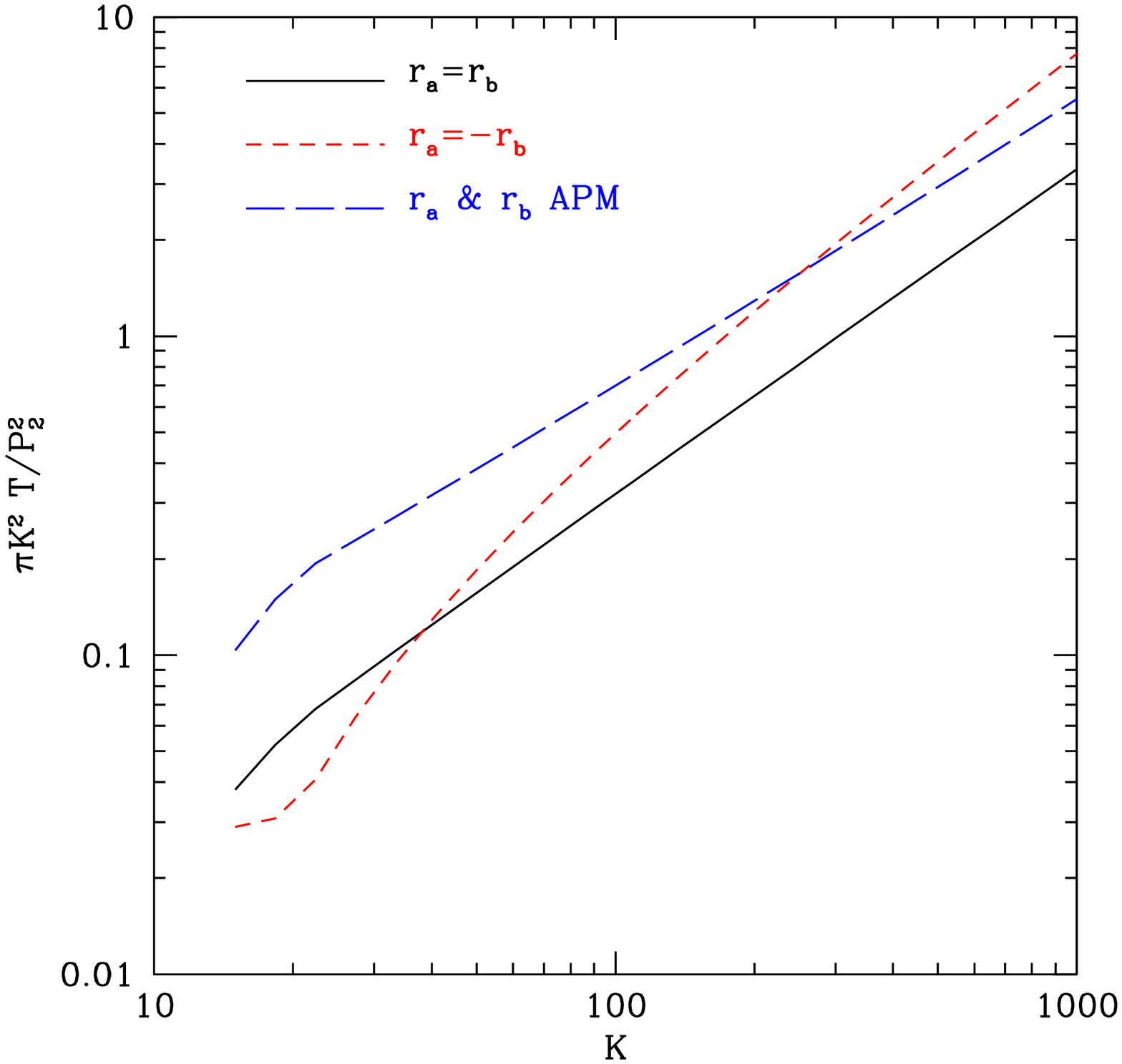}}
\caption{\label{nongratio}
Ratio of the non-Gaussian terms to the Gaussian terms in the diagonal elements
of the covariance matrix of the angular power spectra.  We see that
the non-Gaussian terms are subdominant for $K<100$.}
\end{figure}

In reality, the four-point function has other contributions due to shot
noise, but in the case of APM they are subdominant for the scales of
interest. The full four-point function can be written as
\begin{equation}
\bar T_4^{full}(K,K')={1\over {\bar n}^3}+{2\over {\bar n}^3}
[P_2(K)+P_2(K')]+{\bar B(K,K')
\over {\bar n}} + \bar T_4(K,K')
\end{equation}
where $\bar B$ is the averaged bispectrum over the shells.  
The scaling of the three- and four-point functions with the power
spectrum means that each of the additional terms coming from shot
noise are down by a factor
$P_2(K)\bar n$, which is smaller than one for the measured APM power
spectra up to $K \approx 1000$. 

In Figure \ref{rij}, we show the correlation coefficients for the power
spectra for the different choices of $r_a$ and $r_b$. 
The hierarchical model for the four-point function does not
guarantee that the correlation coefficient stays smaller than
unity, illustrating that this model cannot describe correctly the
correlations induced by gravity.  Only the case $r_a=-r_b$
makes the coefficients stay smaller than one, but \citet{Sco99}
show that the shape of the correlation
coefficients in the simulations are not particularly well fitted by
this choice. The hierarchical model does give a good estimate
of the order of magnitude of the correlations but cannot account for
their shape; this can also be seen in the results of \citet{Mei99}. 
In summary, our calculation of the non-Gaussian effects should be taken 
as an order of magnitude estimate. 

Figure \ref{nongratio}
shows the ratio of the Gaussian to the non-Gaussian terms in the
covariance of the angular power spectrum.  We conclude that the two
contributions are about equal at $K=100$, corresponding approximately
to $1^\circ$. On the $10^\circ$ scale, we expect the inclusion of the 
four-point function in $C_w$ to alter the error bars by less than 10\%. 

A quick estimate of the effect of four-point terms on the
errors of the correlation function can be obtained
using a simple approximation. The four-point function scales as
$P_2^3$, so we approximate $T_4(K,K')$ as $R {\bar P_2} P_2(K)
P_2(K')$, where we have introduced a mean power ${\bar P_2}$. With
this simplification, the non-Gaussian term in $C_w(\theta,\theta')$ is 
$A_\Omega^{-1}R{\bar P_2} w(\theta) w(\theta')$.
In other words, we simply add a overall random fluctuation in the
amplitude of the correlation function, $\hat w(\theta)=(1+\epsilon)w(\theta)$,
with an \rms\ amplitude of 
\begin{equation}
\left<\epsilon^2\right>^{1/2}=0.05\left({R\over 12}\ {\bar P_2 \over 2\times
10^{-4}}\ {1\ {\rm sr} \over A_\Omega}\right)^{1/2}.
\end{equation} 
This model reflects the tendency of modes to become extremely 
correlated in the non-linear regime, such that the shape of the
power spectrum or correlation function becomes far better determined
than the amplitude.

Taking $\sqrt{R{\bar P_2}}=0.05$, we add this additional correlation
to $C_w$ and repeat the calculation of the power spectrum.  The errors on $\Gamma$
increase by about 7\%.  If we double the amplitude of the
effects to $\sqrt{R{\bar P_2}}=0.1$, the errors on $\Gamma$ 
increase by about 30\%.  We expect that this amplitude is
an overestimation of the non-Gaussian corrections.
The best-fit power spectra for these two cases yield
fits to the observed $w(\theta)$ with $\chi^2=42$ and $34$, respectively, 
on 32 degrees of freedom.  
Weakening the off-diagonal terms of the non-Gaussian portion of $C_w$,
so as to step away from total correlation between different $\theta$, 
causes a less severe degradation in the constraints on $\Gamma$.
We therefore conclude that non-Gaussianity should have a
relatively mild effect on our analysis of APM.

\subsection{Lower Bound on Large-Scale Constraints from Angular Data}
\label{sec:best}

One can set a lower bound on the errors of an angular survey
by working directly from the angular power spectrum.  For a 
Gaussian random field with angular power spectrum $P_2(K)$, 
the covariance matrix of the angular power spectrum measured over
the full sky is diagonal, with an variance equal
to $2P_2^2(K)/2K\Delta K$ for a bin of width $K$.  
We take the optimistic assumption that a survey of sky coverage of
$A_\Omega$ steradians will retain this variance with a scaling of $4\pi/A_\Omega$.
We can use the transformation in equation (\ref{eq:Plimber}) to 
convert the inverse covariance matrix of the angular power spectrum into
that of the spatial power spectrum.  The element relating
two bins in spatial wavenumber is
\beq
C_P^{-1}(k,k') = dk\,dk'\int{dK\over K}{A_\Omega\over4\pi}{1\over P_2^2(K)}
	f(K/k) f(K/k'),
\eeq
where $f(\ra)$ is the survey projection kernel and the bins
have width $dk$ and $dk'$.
To compare two spatial power spectra that differ by
$\Delta P(k)$, we integrate $C_P^{-1}$ to find $\chi^2$:
\beq
\chi^2 = \int dk \int dk' \Delta P(k) \Delta P(k') C_P^{-1}(k,k').
\eeq
Defining 
\beq
\Delta P_2(K) = {1\over K}\int dk f(K/k) \Delta P(k)
\eeq
as the angular power spectrum corresponding to the projection of
the difference of the spatial power spectra,
we find
\beq\label{eq:bestchi}
\chi^2 = {A_\Omega\over 4\pi} \int dK\,K 
\left(\Delta P_2(K)\over P_2(K)\right)^2.
\eeq

\begin{figure}[tbp]
\centerline{\plotone{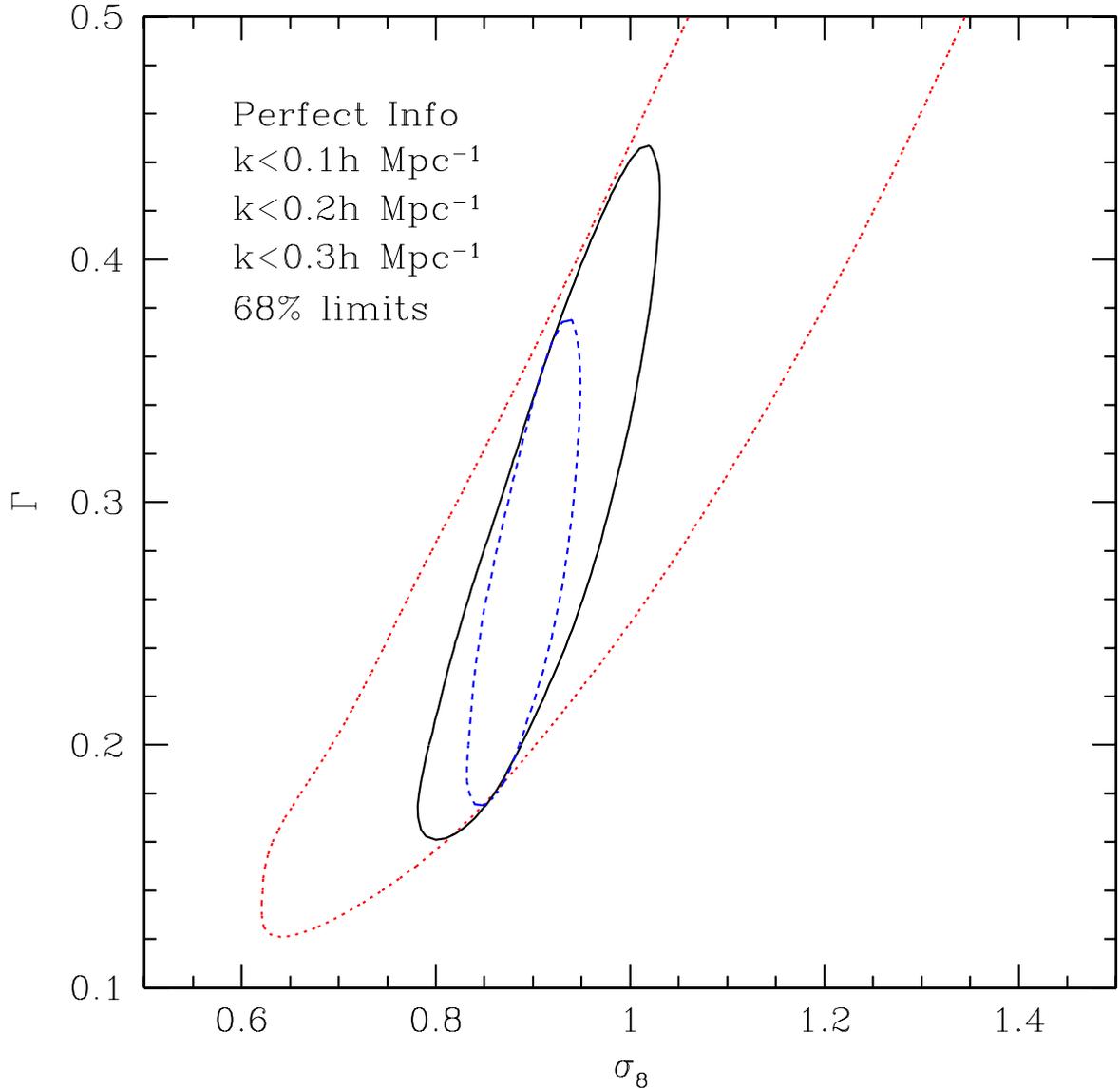}}
\caption{\label{fig:best}
The constraints on CDM parameters within the APM survey if one
adopts the optimistic assumptions of Equation (\protect\ref{eq:bestchi}).
A $\Gamma=0.25$, $\sigma_8=0.89$ model is used to calculate the
sample-variance, and the $\chi^2$ is calculated for the difference
between this model and the grid of other models.
We use non-linear spatial power spectra in all cases.
We compare the CDM models only at scales larger than
$0.1\ihmpc$ ({\it dotted}), $0.2\ihmpc$ ({\it solid}), and 
$0.3\ihmpc$ ({\it dashed}); smaller scales
are assumed to be known perfectly but to contain no extra cosmological
information.
This is the optimistic assumption for the extraction of the
large-scale power spectrum; allowing the small scales to
vary within their errors would worsen the constraints.
We view the regions as lower limits on the uncertainty on the
large-scale power spectrum from APM, save for the minor adjustments that 
would occur with a likelihood analysis on the actual data.
}
\end{figure}

We now apply this limit to the CDM parameter space in the case of APM.  
We assume that the true clustering is given by a $\Gamma=0.25$,
$\sigma_8=0.89$ model with non-linear evolution \citep{Pea96}.  We then
consider how well one can constrain an excursion from this model
on large scales.  We therefore set $\Delta P(k)$ to be zero on scales
$k>k_c$ and equal to the difference between two CDM models (the
model to be tested and the $\Gamma=0.25$ model) on larger
scales.  This corresponds to the limit in which the small scales
are considered to be perfectly known and not allowed to vary within
their errors.  Of course, it also assumes that this perfect 
knowledge on small scales says nothing to distinguish CDM models, 
but we are interested here in the cosmological information available in
the large-scale clustering.  The integral in $K$ is extended from
1 to 1000.

Figure \ref{fig:best} shows the constraints in the $\Gamma$--$\sigma_8$
plane available at scales $k<0.1\ihmpc$, $k<0.2\ihmpc$, and $k<0.3\ihmpc$
using the sky coverage and redshift distribution of the APM survey.
The ranges of allowed $\Gamma$ for the $k<0.2\ihmpc$ case 
are 0.19--0.35 (68\%) and 0.15-0.56 (95\%); the best-fit is $\Gamma=0.25$
by construction.  This is very similar to the limit assigned to the
fit to the power spectrum reconstructed from the actual data if we
use the same CDM model to generate $C_w$.
We conclude that a survey with the sky coverage and selection function
of APM has too much sample variance to place strong constraints on
the shape of the power spectrum on scales greater than $k=0.1\ihmpc$.

While one cannot prove it rigorously, we do not see how one could in practice
achieve errors smaller than the limits 
implied by equation (\ref{eq:bestchi}) and shown in Figure \ref{fig:best}.
The relevant assumptions of Gaussianity, freedom from boundary effects,
infinitesimal bins in angle and wavenumber, total angular coverage,
and perfect information at small scales are all optimistic.
The only subtlety is that equation (\ref{eq:bestchi}) is a statement
about the $\chi^2$ difference between two models, whereas for the
actual survey one is concerned with the likelihood function for model fits
to the data.  This can cause small differences
if the likelihood function is non-Gaussian; in this case, the tendency
would be to shift the allowed region towards larger power, i.e.\ smaller 
$\Gamma$ and larger $\sigma_8$.

\subsection{Comparison to Previous Work}\label{sec:diag}

Despite the limit described in the last section, previous analyses
have found substantially smaller error bars on the large-scale
power spectrum.  In this section, we describe how neglect of
correlations and improper use of smoothing have led to these
underestimates.

We would like to compare the covariance matrix derived from 
theory in \S\ \ref{sec:cov}  to that used in previous analyses
of the power spectra inferred from APM angular clustering.  
Generally, the errors for large-scale correlations have been estimated 
as the deviation between four subsamples of the APM survey 
\citep{Mad96,Bau93}.  
This procedure is at best marginal for estimating even the diagonal 
elements of the covariance matrix, but it is completely inadequate
for estimating the full covariance matrix.  Indeed, one could
only generate four non-zero eigenvalues!  Therefore, the covariance
matrices of either the angular correlations or the spatial power spectra
have been assumed to be diagonal.  Neither of these approximations is
correct or, as we will see, particularly good.

We begin with the angular correlation function.
Without the correlations between bins, it is very easy to overestimate
the power of the data set by using too fine a binning in $\theta$.  
Neighboring bins that are highly correlated will show the same 
dispersion between the subsamples, but one will count this as two
independent measurements rather than one.  The errors on any fit
will improve by $\sqrt{2}$.  The visual cue that this is occurring
is when the subsamples show coherent fluctuations around the mean
rather than rapid bin-to-bin scatter.  This is clearly occurring
in Figure 27 of \citet{Mad96}.

\begin{figure}[tbp]
\centerline{\plotone{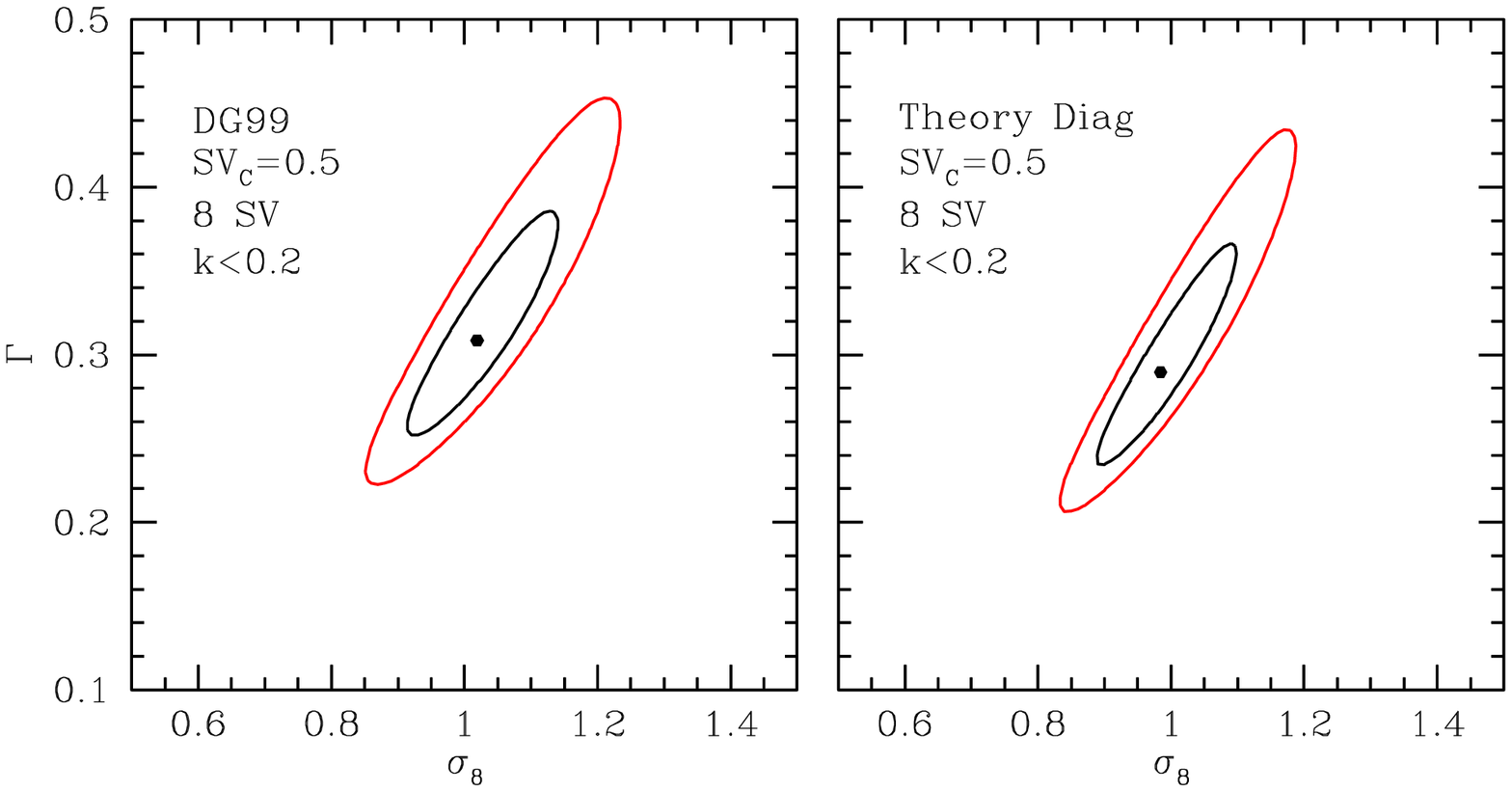}}
\caption{\label{fig:pc4_6}
Constraints when the covariance matrix $C_w$ is assumed to be diagonal.
({\it left panel}) Observed APM error bars \protect\citep{Mad96,Dod99} 
formed into a diagonal covariance matrix.
({\it right panel}) Covariance matrix from \S\ \protect\ref{sec:cov}, but
with the off-diagonal terms set to zero.
In each case, we use $SV_C=0.5$, consider only the largest 7 SV when
constructing $\bfP'$, and use only wavenumbers $k<0.2\ihmpc$ in the fit.
These constraints should be compared to those of Figure 
\protect\ref{fig:pc1_1}.
}
\end{figure}

We can compare our calculation of $C_w$ to the quoted observational errors 
by setting all of our off-diagonal terms to zero.
We then substitute this new $C_w$ 
and recalculate limits on $\Gamma$ and amplitude
in the manner described in the previous section.  
We do the same for a diagonal $C_w$ that uses the errors 
on $w(\theta)$ based on the dispersion between 4 subsamples of APM
\citep{Mad96,Dod99}.
As shown in 
Figure \ref{fig:pc4_6}, these two choices give constraint regions 
that are quite similar to one another.  
The fact that these two treatments give similar results is
evidence that sample variance in the Gaussian limit does
explain most of the observed scatter in $w(\theta)$ on large
angular scales in APM and further justifies the approximations
that underlie our estimation of $C_w$.

Importantly, both diagonal treatments give constraints on CDM parameters
that are a factor of two tighter than those found when
using the theoretical covariance matrix with its off-diagonal
terms.  For example, comparing the 68\% semi-range on $\Gamma$, we find
0.09 in the full $C_w$ case and 0.043 in either of the diagonal 
counterparts.  The best-fit $\Gamma$ in the diagonal cases are
around 0.3, somewhat higher than in the analysis with non-zero
correlations in $w(\theta)$ and suggesting a bias in the reconstruction.

It should also be noted that when either of these diagonal covariance
matrices are used, the $\chi^2$ for the $w(\theta)$ of the best-fit
power spectra is less than 3 for 32 degrees of freedom.  This is 
another indication that these matrices do not properly describe the
error properties of the data.

DG99 reconstruct the power spectrum based on a diagonal covariance 
matrix.  However, they obtain limits using $k<0.124\ihmpc$ that are 
tighter than what we show for $k<0.2\ihmpc$ in Figure \ref{fig:pc4_6}.
We believe that this is caused by the way in which their smoothing
prior enters the calculation of the covariance matrix,
namely that the quoted covariance matrix is for the smoothed estimator 
of the power not the actual power itself.  
On scales where the constraints from the data are poor, the smoothing
prior will choose a value of the power based on an extrapolation from
the wavenumbers with stronger measures of the power.  The variance
between samples of 
this extrapolated power will be far smaller than the true uncertainty
in the power.
We think that this underestimate of the errors at large scales is 
responsible for the discrepancy between the SVD treatment and the 
method of DG99.  Indeed, DG99 found that the errors on cosmological
parameters increase as they relaxed the smoothing prior.

Baugh \& Efstathiou (1993, 1994) did not use the covariance on $C_w$; instead
they estimated errors on $P(k)$ directly by using the variance of the
power spectra of the 4 subsamples, having inverted each separately.
Again, 4 subsamples was too few to estimate the off-diagonal terms of
$C_P$.  It is clear, however, that these terms are important, as
Figure 9 of BE93 reveals that the $P(k)$ from the 4 subsamples do show 
obvious correlations in their differences from the mean.  The
tests on simulations in \citet{Gaz98} also neglect the correlations
between different bins in the reconstructed power spectra.

Unfortunately, we cannot simply use the diagonal terms 
of our $C_P$ matrix to compare to the BE93 results, because  
the inversion procedure of BE93 includes an implicit smoothing prescription.
In order to compare to their estimate of the errors on the smoothed power
spectrum, we would need to project our $C_P$ matrix onto their 
allowed basis, removing the variance in any disallowed directions
in $P$-space.  Without this step, the large variances we included
for the ill-constrained directions will give enormous variance to
individual $k$ bins when the detailed correlations between bins
are discarded.

One must be especially wary of estimating error bars from the variance
between subsamples when a smoothing prior has been applied to a 
wavenumber or angle where the data is not constraining.  In the present 
context, the Lucy inversion method employed by BE93 contained a 
smoothing step that pushed $P(k)$ to a particular functional form.  
When such a method is used on large scales where the power is 
not well-constrained, 
then all subsamples will tend to reconstruct a power spectrum 
value on large scales that is simply an extrapolation of the 
smaller scale result.
The dispersion between the subsamples will not grow with scale
as fast as they would in the absence of smoothing,
causing the resulting error bars to be significantly underestimated 
on large scales.  We suspect that this effect is a significant 
contribution to why \citet{Bau93} find near constant power and
small errors at $k<0.05\ihmpc$.

\section{Conclusion}\label{sec:concl}

Both the angular correlation function and the spatial power spectrum
inferred from angular clustering have important correlations between
different bins of angle and wavenumber even if the fluctuations
are Gaussian.  Previous analyses of the deprojection of angular
clustering have neglected these effects.
In this paper, we have shown how to include sample variance in the 
covariance matrix
for the angular correlation function $w(\theta)$ under a Gaussian, wide-field
approximation.  We have then described how one may invert $w(\theta)$
to find the spatial power spectrum using singular value decomposition
in such a way as to retain the full covariance matrix.
The method allows one to handle the near-singularity of
the projection kernel without numerical difficulty and can yield
a smoothed version of the deprojected power spectrum 
without sacrificing the covariances of unsmoothed spectrum.

Using the large-angle galaxy correlations of the APM survey as an
example, we have shown that correlations between different bins in $\theta$
and in $k$ are critical for quoting accurate statistical limits on
the power spectrum and model fits thereto.  With the sample variance
properly included, we find that APM does not detect a downturn in
$P(k)$ at $k<0.04\ihmpc$; the significance is only 1--$\sigma$.  
Fitting non-linearly extrapolated, scale-invariant CDM power spectra 
to the power spectrum at large scales ($k<0.2\ihmpc$), we find 
that APM constrains the CDM parameter $\Gamma$ to be 0.19--0.37 (68\%).
We have investigated a wide range of alterations to the method in
the hopes of shrinking this range but have found nothing that
makes a significant difference.
Indeed, in \S\ \ref{sec:best}, we showed that the above 
constraints already approach the best available to a survey
with the sky coverage and selection function of APM.
Extending the CDM fits to smaller scales would improve the constraints,
but this depends entirely on the modeling of galaxy bias and non-linear
gravitational evolution.  Moreover, such a fit wouldn't validate this
particular set of CDM models because many other models would look 
similar in the non-linear regime.  To confirm a model from galaxy
clustering, one would like to see the characteristic features of the
model directly rather than attempt to leverage a measurement of 
the slope in the non-linear regime onto a cosmological parameter
space.

We have made a number of approximations in our analysis.  In our
treatment of large scales, we have ignored the ability of boundaries
to alias power from one scale to another and used Limber's
equation even for modes with wavelengths similar to the scale of the 
survey.  On small scales, we have ignored three-point and four-point
contributions to the covariance matrix of the angular statistics.
In general, we have treated the likelihood of the correlation
function as a Gaussian and ignored the fine details of how cosmology
or evolution of clustering might enter.  We have argued that the above
approximations are likely to be reasonably accurate for an analysis
of the large-angle clustering of APM.  We also have not questioned
the redshift distribution function that has been used in past APM
analyses nor included any systematic errors.  
Conservatively, therefore, one can regard our results as the optimistic
limits, because it is very unlikely that the breakdown of any of the
above assumptions would actually improve the constraints!

Surveys such as DPOSS and SDSS will be substantially deeper and wider
than the APM survey.  We repeat the analysis of \S\ \ref{sec:best}
for parameters suggestive of SDSS, namely 3.1 steradians of sky coverage
to a median redshift of 0.35.  This yields an error on $\Gamma$ of
0.017 (1--$\sigma$) about a fiducial model of $\Gamma=0.25$.  
Remember that this is an optimistic limit on the error, that
we have assumed perfect knowledge of the selection function, and that
we have only allowed one other parameter, the amplitude, to vary!
With analogous assumptions, the limit on $\Gamma$ from the SDSS redshift
survey of bright red galaxies is roughly 0.007.  The
redshift survey would be more strongly preferred if one were interested in
narrower features in the power spectrum \citep{Mei99b}, such
as would be needed to separate effects in a larger parameter
space \citep{Eis99}.  As regards systematic errors, the angular
survey suffers from its dependence on purely tangential modes,
while the redshift survey must contend with redshift-space distortions.

If one could use color information to select a clean,
high-redshift sample of galaxies, the prospects for interpreting 
angular correlations improve.  For example, using only those galaxies
with $z>0.45$ in the above SDSS example
drops the limiting errors on $\Gamma$ to 0.010.  This occurs because
the obscuring effects of smaller-scale clustering from lower redshift
galaxies have been removed; moreover, the projection from three dimensions
to two becomes significantly sharper.  This performance is comparable
to that of the redshift survey and would allow measurement of the 
large-scale power spectrum in a range of redshifts disjoint from
the spectroscopic survey, thereby
allowing one to study the evolution of large-scale clustering.

The results of this paper, and particularly the limits set in 
\S\ \ref{sec:best}, provide a cautionary note for the interpretation
of large angular surveys.
Even at the depths of the SDSS imaging, the constraints on large-scale 
clustering from angular correlations alone are not strong.  The inclusion 
of photometric redshifts to separate the sample into multiple
(or even continuous) radial shells could provide a significant
improvement to this state of affairs.  

\acknowledgments We thank Scott Dodelson, Enrique Gazta\~naga, 
Lloyd Knox, Jon Loveday,
Roman Scoccimarro,
Istvan Szapudi, Max Tegmark, and Idit Zehavi for helpful discussions.
Support for this work was provided by NASA through Hubble
Fellowship grants HF-01118.01-99A (D.J.E.) and HF-01116.01.98A (M.Z.)
from the Space Telescope Science
Institute, which is operated by the Association of Universities
for Research in Astronomy, Inc, under NASA contract NAS5-26555.
D.J.E.\ was additionally supported by Frank \& Peggy Taplin Membership 
at the IAS.

\end{document}